\documentclass[12pt]{article}
\usepackage{array}
\usepackage{amsmath}
\usepackage[margin=2.0cm]{geometry}
\usepackage{epsfig}
\usepackage{cite}
\usepackage{amsfonts}
\usepackage{fancyhdr}
\usepackage{setspace}
\usepackage{enumitem}
\usepackage{threeparttable}
\usepackage{graphicx}
\usepackage[font=small,skip=4pt]{caption}
\usepackage{subcaption}
\usepackage[font={sf,sl}]{caption}
\usepackage{bm}
\usepackage{booktabs}
\usepackage{tablefootnote}
\usepackage{cancel}
\usepackage{wasysym}
\usepackage{amssymb}
\usepackage{float}
\usepackage{lineno}
\usepackage{algorithm}
\usepackage{algpseudocode}
\usepackage{microtype}
\usepackage{wrapfig}
\usepackage{xcolor}
\usepackage{textcomp}
\usepackage{arydshln}
\usepackage{tikz}
\usepackage{authblk}
\usepackage{titlesec}
\usepackage{hyperref}
\usepackage[export]{adjustbox}
\usepackage[page,toc,titletoc,title]{appendix}
\usetikzlibrary{arrows}
\newcommand{\vx}{\mathbf{x}}
\newcommand{\vn}{\mathbf{n}}
\newcommand{\rspear}{\rho_{\mathrm{s}}}
\titleformat{\section}
{\normalfont\bfseries\scshape}{\thesection}{1em}{}

\titleformat{\subsection}
{\normalfont\bfseries\scshape}{\thesubsection}{1em}{}

\titleformat{\subsubsection}
{\normalfont\slshape}{\thesubsubsection}{1em}{}

\titleformat{\title}{\normalfont\bfseries}{\thesection}{1em}{}

\title{\vspace{-3em} \large \bfseries Uncertainty-guided active learning for surrogate prediction of stream-finishing wear fields.}

\author[1]{\normalsize Anand Kumar}
\author[2]{\normalsize Puli Saikiran}
\author[2]{\normalsize Vineet Dawara}
\author[2]{\normalsize Koushik Viswanathan\thanks{koushik@iisc.ac.in}}
\affil[1]{Dept. of Design \& Manufacturing, Indian Institute of Science, Bangalore}
\affil[2]{Dept. of Mechanical Engineering, Indian Institute of Science, Bangalore}

\date{\vspace{-1em}\normalsize \textsc{\today}}
\numberwithin{equation}{section}
\numberwithin{table}{section}
\begin{document}

\maketitle
\thispagestyle{plain}
\vspace{-1em}
\hrulefill

\begin{abstract}
In stream finishing, the wear experienced by a workpiece depends strongly on its orientation within the rotating abrasive media. Determining suitable orientations to achieve uniform wear requires evaluating the wear-rate field over all feasible orientations. Although the discrete element method (DEM) accurately resolves particle interactions, simulating hundreds of feasible orientations for a new geometry is computationally expensive. We present an uncertainty-guided surrogate framework that predicts, directly from geometry, the three fields governing erosion: per-triangle normal impact velocity, tangential impact velocity, and particle impact flux. These fields are combined through the Finnie wear model to reconstruct the wear-rate distribution. The surrogate employs a deep ensemble whose disagreement estimates epistemic uncertainty, enabling an active-learning strategy that selectively performs DEM simulations for the most uncertain orientations. Trained using only $13\%$ of the $696$ feasible orientations, the surrogate achieves Spearman rank correlations of $0.93$, $0.89$, and $0.93$ for the normal impact velocity, tangential impact velocity, and particle impact flux, respectively. Moreover, the predicted uncertainty is well calibrated, reliably anticipating prediction error and the fidelity of the reconstructed wear field, which matches DEM with a Spearman rank correlation of up to $0.97$ for low-uncertainty orientations and degrades in a controlled manner as uncertainty increases.
\end{abstract}

\section{Introduction}\label{sec:intro}

Stream finishing is an emerging abrasive finishing process in which a workpiece is held stationary within a rotating container filled with abrasive granular media. The relative motion between individual granules and the workpiece surface results in material removal and, therefore, surface finish enhancement \cite{ref_sf1}. This process is particularly well-suited for precision and complex-shaped components, such as turbine blades, orthopedic implants, and sculptured surfaces. For these intricate geometries, conventional mass finishing processes—like barrel tumbling and centrifugal finishing—often struggle to achieve a uniform finish due to the unequal exposure of abrasive particles across different surface regions \cite{ref_sf_hist}.A decisive process variable in stream finishing is the orientation of the workpiece relative to the media flow. Different orientations expose distinct surface regions to the abrasive stream, thereby producing different spatial wear-rate distributions. To predict these distributions and identify suitable processing orientations, it is very common to use approximate continuum flow models with a suitable constitutive equation (e.g., the $\mu(I)$ model) with finite-element solvers \cite{ma2023numerical}. However, the predictive accuracy of such models are often quite questionable,  when applied to complex geometries. 

Consequently, the Discrete Element Method (DEM) \cite{ref_dem1} has emerged as the most robust and accurate computational approach for predicting wear rates across varying orientations. When coupled with a microscopic erosion model, such as the Finnie model \cite{ref_finnie}, DEM provides high-fidelity predictions of localized particle–surface interactions and resulting spatially varying material removal \cite{ref_dem1, ref_dem2, zhao2026agent}. DEM predictions of finishing wear have been validated against experiments in several studies~\cite{ref_dem_val, roessler2023calibrated, aydin2025simulation}, making it a very appealing method for predicting wear-rate patterns. DEM simulations have also been used recently to establish preliminary process parameters in the context of stream finishing of simple geometries \cite{zanger2019optimization}. 

However, in practice, especially with complex part contours, uniform or targeted finish is rarely achieved through a single orientation alone. This requires a constructed sequence of orientations, each contributing to material removal in specific regions. Designing such a sequence for a given geometry requires mapping the spatial wear-rate distribution for every feasible orientation. Under typical experimental work fixturing constraints, discretizing the orientation space can often generate a very large number of candidate orientations. As an example, rotating a non-symmetric workpiece about the $x$, $y$, and $z$ axes through a pivot point in relatively coarse increments of $15^{\circ}$, generates more than a thousand feasible configurations. Given its computational cost, conducting an exhaustive DEM simulation programme for every orientation of a new geometry is hence practically prohibitive. This limitation must be circumvented if DEM is to be used for creating part orientation strategies based on wear rate distributions.

A surrogate model that predicts erosion behavior for any orientation from a small number of DEM simulations would relieve this computational cost, and a deep neural network is a natural candidate for such a surrogate. Simulation-generated abrasive-based finishing process surface images, combined with transfer learning, are data-driven surrogates used to predict surface roughness without exhaustive experimental sampling ~\cite{yi2024grinding}. Neural networks, however, require large datasets, and training one to the accuracy required here would ordinarily demand far more simulated orientations than is often practically feasible. What is needed is therefore a predictor that also reports on its inherent uncertainity, so that DEM effort can be directed to precisely the orientations where the model is least reliable, and the training set can be grown economically rather than exhaustively. The most common approach to this class of problem is Bayesian optimization, also known as efficient global optimization~\cite{jones1998efficient, yang2022bayesian, chae2010kriging, kanazaki2007efficient}. Briefly, it reduces model uncertainty by iteratively updating the surrogate according to an acquisition function~\cite{snoek2012practical, shin2020bayesian, shimoyama2013kriging} that combines the predicted value with its uncertainty to decide where to sample next. This active, uncertainty-guided sampling method is what we adopt here. 

Several approaches exist to a neural network with such uncertainty estimates. Bayesian neural networks place a distribution over the weights and infer a posterior, but are computationally expensive and difficult to scale~\cite{blundell2015weight}. Monte Carlo dropout offers a cheaper approximation by sampling dropout masks at inference~\cite{gal2016dropout}, but this technique has been reported to be overconfident on complex networks. Deep ensembles, in which several independently initialized networks are trained and their disagreement taken as the uncertainty, are simpler to implement and have been shown to give better-calibrated uncertainty than both alternatives, particularly on out-of-distribution inputs~\cite{ref_ensembles, ovadia2019can, fort2019deep}. We therefore adopt a deep-ensemble surrogate, using its epistemic uncertainty as the acquisition signal in place of the Gaussian-process posterior variance conventionally used in Bayesian optimization.

This paper is organized as follows. Section~\ref{sec:methods} presents the methodology in two parts: the DEM model of the stream-finishing process, including the feasible orientation set and the Finnie wear relation (Section~\ref{sec:dem_setup}), and the uncertainty-aware surrogate itself, covering the deep-ensemble architecture, the physics-informed input features, the seed-selection procedure, and the execution of the active-learning loop (Section~\ref{sec:surrogate}). The DEM model serves both as the source of training data for the surrogate and as the  ground truth against which its predictions and its uncertainty are validated. Section~\ref{sec:results} first establishes that the predicted uncertainty is truthful, then reports the accuracy of the velocity and granule flux predictions across iterations, and ends with wear fields reconstructed through the Finnie model.

\section{Methods}\label{sec:methods}
\subsection{DEM simulation of stream finishing and problem formulation}
\label{sec:dem_setup}
The discrete element method (DEM), introduced by Cundall and Strack~\cite{ref_dem1}, resolves granular flow by tracking every particle individually and integrating Newton's equations of motion for each. Inter-particle and particle--wall interactions are treated as elastic contacts, with normal contacts following the nonlinear Hertzian spring--dashpot formulation and tangential contacts the Mindlin--Deresiewicz model~\cite{mindlin1953elastic}. DEM resolves each collision explicitly and therefore records the position, velocity, and surface facet of every impact, the level of detail required for stream finishing, where material removal is driven by individual abrasive impacts. This resolution is also responsible for the method's cost, as simulation stability requires a time step smaller than the contact duration, which results in a single orientation taking hours of computation, and mapping the wear field over a large orientation set by DEM alone is impractical.

\subsubsection{Simulation setup}\label{sec:sim_setup}
\begin{figure}[t]
\includegraphics[width=\textwidth]{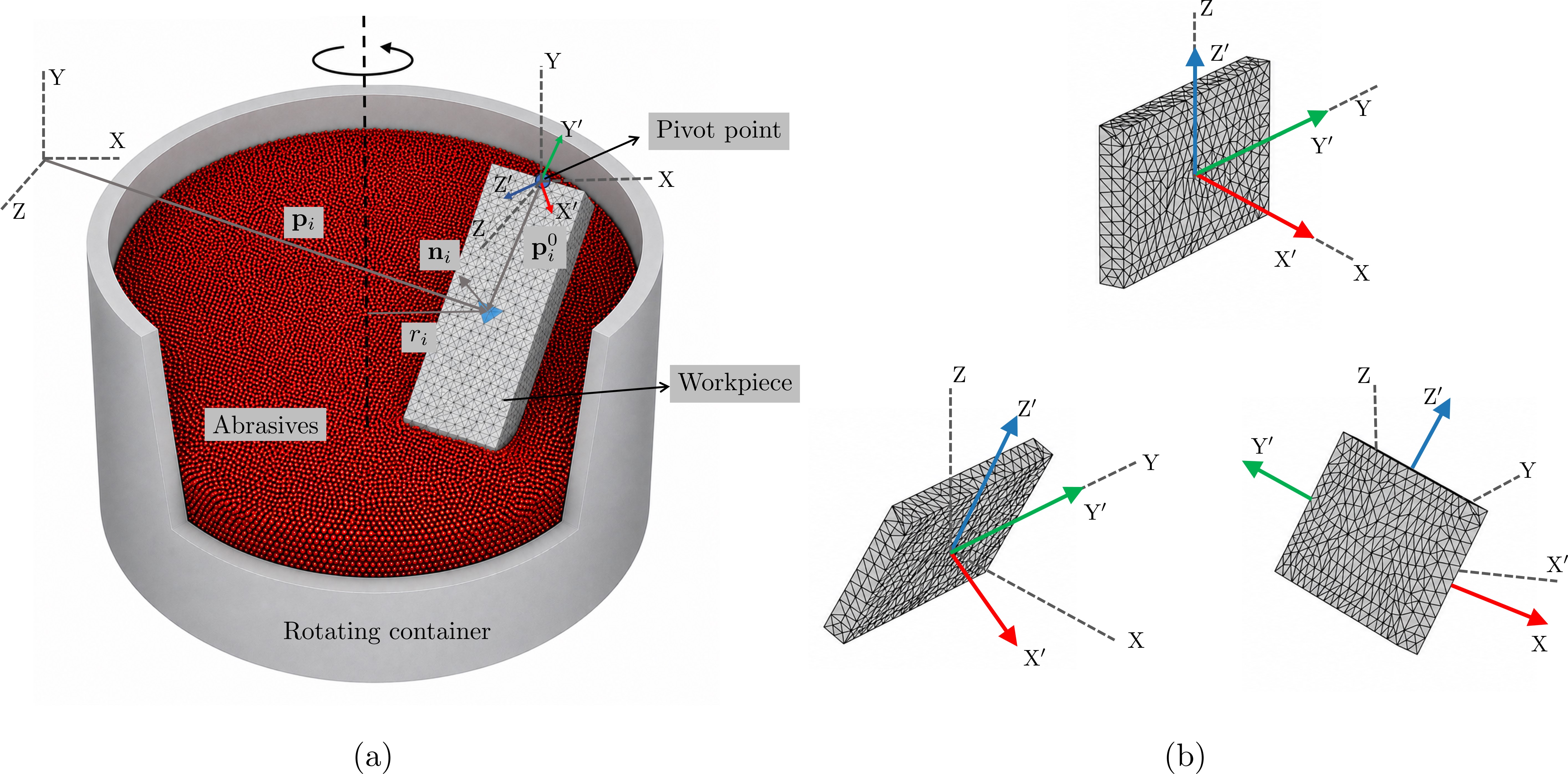}
\caption{Computational setup and workpiece orientations. (a) The cylindrical container holds the abrasive media and rotates about its vertical axis. The workpiece is immersed in the media bed so that the circulating abrasive flow impacts its surface. One end of the workpiece is defined as the pivot point about which all orientations are prescribed. (b) Examples of orientations obtained by rotating the baseline orientation about the global axes: \text{XYZ 0 0 0}, \text{XYZ 0 30 0} ($30^\circ$ about $Y$), and \text{XYZ 0 30 60} ($30^\circ$ about $Y$ then $60^\circ$ about $Z$). Coloured arrows denote the rotated body axes $X', Y', Z'$, and dashed grey lines denote the fixed global axes.
}
\label{fig:setup_orientation}
\end{figure}

We use a commercial solver \textit{Ansys Rocky 2024R2} for the DEM simulations. The setup, shown in Fig.~\ref{fig:setup_orientation}(a), consists of a hollow cylindrical container of diameter $150$~mm and height $100$~mm filled with spherical silicon carbide abrasive particles of diameter $2$~mm. A cuboidal workpiece (assumed to be stainless steel 304 grade) with dimensions $60 \times 35 \times 7$~mm$^3$ is clamped from the side of the pivot point shown in Fig.~\ref{fig:setup_orientation}(a) by a fixture and immersed in the media bed in a prescribed orientation. These dimensions are chosen to coincide with a separate experimental setup. Detailed comparisons between the numerics and experimentally observed wear patterns are outside the scope of the present work and not discussed further. We hope to present them in a future manuscript. 
The container rotates at $120$~rpm about its vertical axis, setting the media into a circulating stream that flows past the immersed workpiece and strikes its surface, and these repeated particle impacts remove the material. A total of $1.98\times10^{5}$ particles are used, corresponding to a packing fraction of approximately $0.55$. To increase the stable integration time step, the Young's modulus of the particles and container walls is reduced by a factor of $10^{4}$, which has been shown in prior work to preserve bulk flow behavior~\cite{lommen2014speedup}. The DEM results are accordingly interpreted in terms of relative wear patterns and spatial wear localization rather than absolute wear magnitudes. All parameters used for the simulations are summarized in Table~\ref{tab:dem_params}.

\begin{table}[t]
\centering
\caption{Simulation parameters used in the DEM model of the stream-finishing process.}
\label{tab:dem_params}
\begin{tabular}{lll}
\toprule
\textbf{Parameter} & \textbf{Symbol} & \textbf{Value}\\
\midrule
\multicolumn{3}{l}{\textit{Container}}\\
Container diameter          & $D_c$   & $150$ mm\\
Container height            & $H_c$   & $100$ mm\\
Rotational speed            & $\omega$ & $120$ rpm\\
Packing fraction            & $\phi$      & $0.55$\\
\midrule
\multicolumn{3}{l}{\textit{Abrasive media}}\\
Material                    & --      & Silicon carbide (SiC)\\
Particle size               & $d_p$   & $2.0$ mm\\
Particle shape              & --      & Spherical\\
Particle density            & $\rho_p$ & $3200$ kg\,m$^{-3}$\\
Number of particles         & $N_p$   & $1.98\times10^{5}$\\
\midrule
\multicolumn{3}{l}{\textit{Workpiece}}\\
Material                    & --      & SS304 stainless steel\\
Dimensions                  & --      & $60\times35\times7$ mm$^3$\\
Surface triangles           & $N$     & $1180$\\
\midrule
\multicolumn{3}{l}{\textit{Contact parameters}}\\
Young's modulus, particle    & $E_p$   & $2.0\times10^{8}$ Pa\\
Young's modulus, workpiece   & $E_s$   & $2.1\times10^{7}$ Pa\\
Coefficient of restitution   & $e$     & $0.32$\\
Friction, particle--surface  & $\mu_{ps}$ & $0.48$\\
Friction, particle--particle & $\mu_{pp}$ & $0.35$\\
Poisson's ratio              & $\nu$   & $0.3$\\
\bottomrule
\end{tabular}
\end{table}

\subsubsection{Workpiece orientation and the feasible orientation set}\label{sec:orientation}
The severity of particle impingement is, as is to be expected, not uniform over the entire workpiece surface. Facets that face into the incoming stream are understandably struck more often and at higher speed than facets that sit in the wake of the workpiece, so the wear pattern depends on how the workpiece is oriented in the flow. An orientation is parametrised by the triplet $\bm{\theta} = (\alpha_x, \alpha_y, \alpha_z)$ of rotation angles about the $x$, $y$, and $z$ axes, applied sequentially about a fixed pivot at the top of the workpiece where it is clamped by the fixture, and is labelled \text{XYZ} $\alpha_x\,\alpha_y\,\alpha_z$. These two coordinate systems are shown schematically in Fig.~\ref{fig:setup_orientation}, along with the construction with three examples (see panel (b)). These correspond to baseline \text{XYZ 0 0 0}, a single rotation of $30^{\circ}$ about the $y$ axis giving \text{XYZ 0 30 0}, and a compound rotation of $30^{\circ}$ about $y$ followed by $60^{\circ}$ about $z$ giving \text{XYZ 0 30 60}. Discretizing each angle in $15^{\circ}$ steps, and removing configurations that require holding at other locations than the pivot point (experimental constraint) and restricting to only two-angle rotations, leaves a feasible set $\mathcal{F}$ of $696$ orientations, which is the design space considered in the rest of the manuscript.

\subsubsection{Per-triangle impact data and the Finnie wear model}\label{sec:finnie}
The workpiece surface is discretized into a triangulated mesh of $N=1180$ elements, and the quantities presented (e.g., velocity) are resolved at the level of these individual triangles. For every collision between a granular particle and the workpiece surface, the DEM solver records the impact position, impact velocity, and the surface triangle/facet corresponding. Averaging these per-collision records over a steady-state time window, once the media flow has settled into a statistically stationary state, reduces them to separate per-triangle fields. These are the mean normal impact velocity $v_n$, mean tangential impact velocity $v_t$, and particle impact flux $\phi$ (impacts per unit time). Erosive wear is related to these primary fields through the Finnie model for ductile materials~\cite{ref_finnie},
\begin{equation}
\dot{w} \;=\; k\,\phi\,V^{2} f(\gamma),
\qquad
V = \sqrt{v_n^2+v_t^2},
\qquad
\gamma = \arctan(v_n/v_t),
\label{eq:finnie_model}
\end{equation}
where $f(\gamma)$ is Finnie's piecewise angular function with its maximum near $\gamma_0 = 18.5^{\circ}$ and $k$ is a material constant. The essential point for what follows is that if the velocity components and the flux are known for a triangle, its wear rate follows analytically from Eq.~\ref{eq:finnie_model}. Resolving the wear field of an orientation, therefore, reduces to resolving these three primary fields. A single DEM simulation provides them for one orientation only. To obtain a sequence of orientations for our surface-finishing problem, we need this information for all $696$ feasible orientations, and since DEM is computationally intensive, this is a challenging task. This necessitates an alternative data-driven surrogate model that can efficiently predict the wear rate of feasible orientations.
\subsection{Uncertainty-aware surrogate model}\label{sec:surrogate}
\subsubsection{Prediction of kinematic fields \emph{in lieu} of direct wear-rate estimates}\label{sec:why_primary}
We note that a surrogate could regress the wear rate obtained from each DEM simulation directly. We instead predict the primary fields and reconstruct wear through Eq.~\ref{eq:finnie_model}. The reason is the structure of the Finnie relation. The quadratic velocity dependence and the sharply peaked angular function $f(\gamma)$ make the wear rate a strongly non-linear, non-smooth function of orientation, whereas the velocity components and the flux vary smoothly. Predicting smooth quantities and confining the sharp non-linearity to the analytical law is both easier to learn and physically transparent.

\subsubsection{Deep ensemble and epistemic uncertainty}\label{sec:ensemble_unc}
\begin{figure}[t]
\centering
\includegraphics[width=\textwidth]{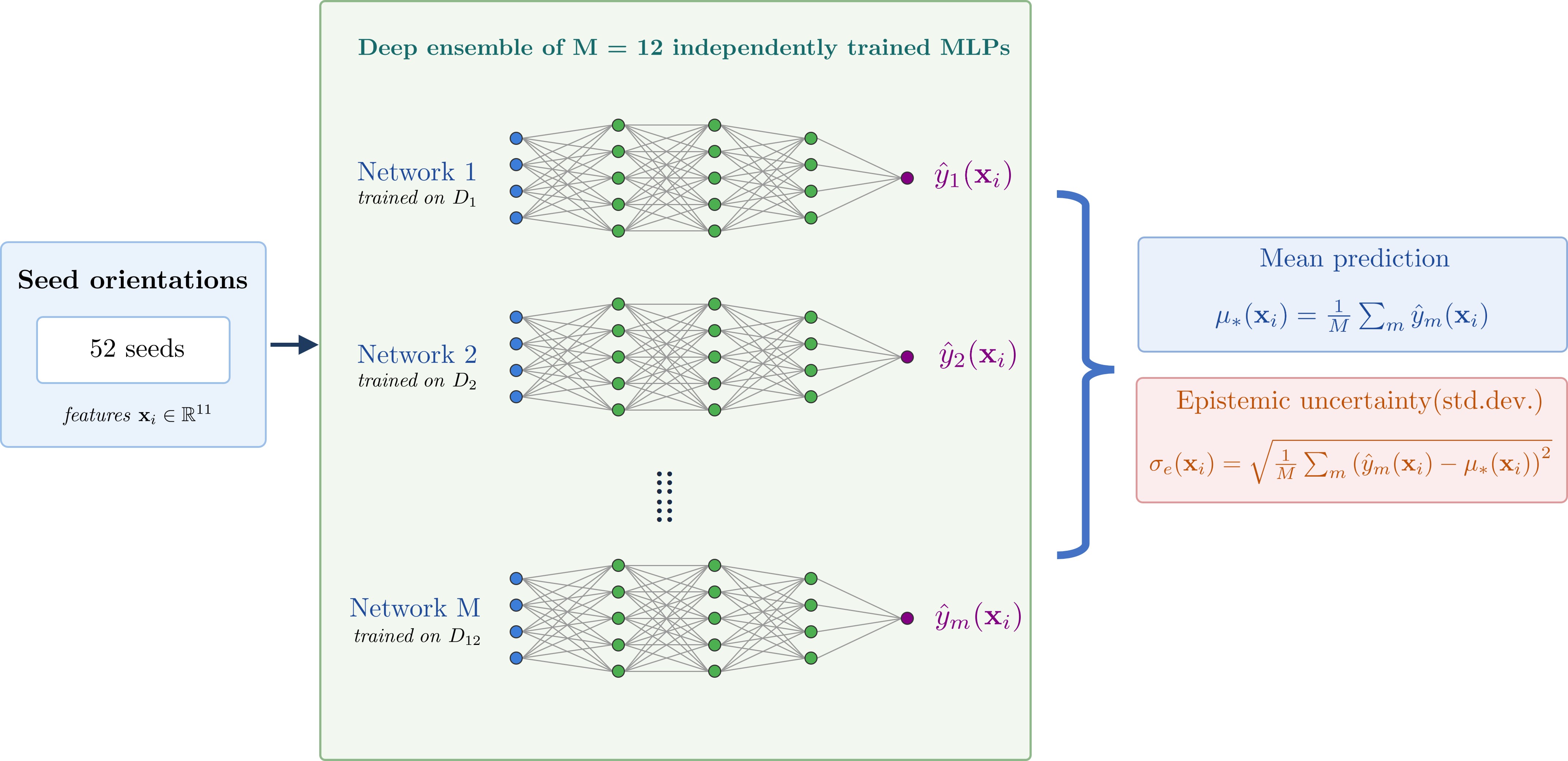}
\caption{Deep-ensemble surrogate. Each of the $M=12$ networks shares the same architecture but starts from a different random initialization and is trained on a different random $80\%$ subset $D_m$ of the seed orientations. For a given input, the $M$ member predictions $\hat{y}_1,\dots,\hat{y}_M$ are combined into the ensemble mean $\mu_*$ (the prediction used) and the between-member standard deviation $\sigma_e$ (the epistemic uncertainty).}
\label{fig:nn_framework}
\end{figure}

\begin{figure}[t]
\centering
\includegraphics[width=0.6\textwidth]{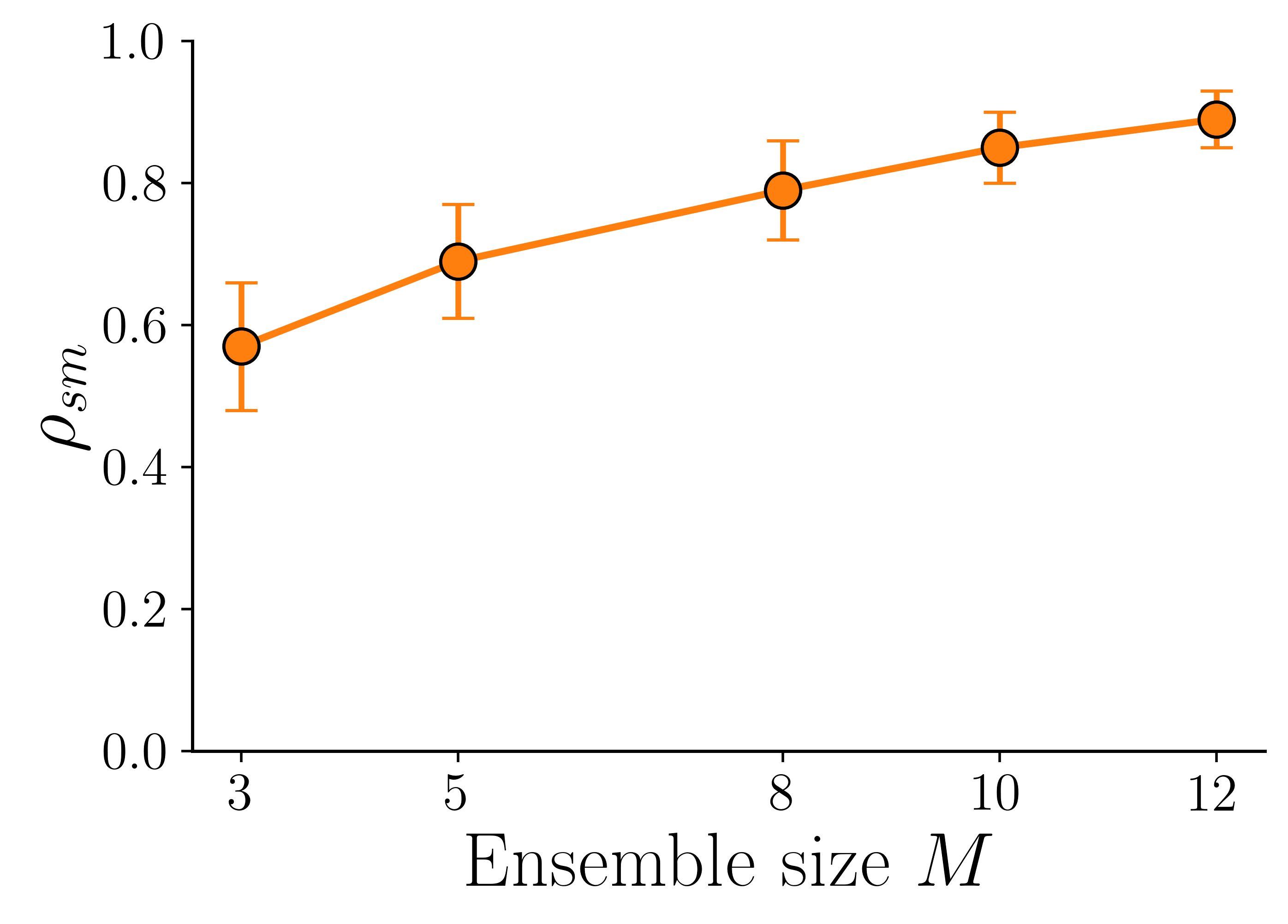}
\caption{Sensitivity of the orientation-level epistemic-uncertainty ranking to the ensemble size $M$. The Spearman rank correlation $\rho_{sm}$ quantifies the stability of the uncertainty-based ranking as $M$ is varied; error bars denote the spread across repeated ensemble realisations. The ranking stabilises as $M$ increases, with diminishing returns beyond $M=10$, supporting the choice $M=12$ used throughout this work.}
\label{fig:ensemble_size}
\end{figure}

The surrogate must report how much a prediction can be trusted, not merely the prediction itself, because the decision of where to spend further DEM effort rests on it. The uncertainty that matters for this decision is the epistemic (model) uncertainty, which arises because the finite training set does not completely constrain the fitted model, and it reduces when informative data are added. A deep ensemble~\cite{ref_ensembles} estimates this quantity directly. As illustrated in Figure~\ref{fig:nn_framework}, $M$ networks of identical architecture are trained independently. Diversity among ensemble members is induced in two ways: each network starts from a different random weight initialization, and is trained on a different random $80\%$ subset $D_m$ of the available training orientations. For an input $\vx$, member $m$ returns a prediction $\hat{y}_m(\vx)$, and the ensemble prediction and its epistemic uncertainty are
\begin{align}
\mu_*(\vx)_i &= \frac{1}{M}\sum_{m=1}^{M}\hat{y}_m(\vx)_i,
\label{eq:ens_mean}\\
\sigma_e^2(\vx)_i &= \frac{1}{M}\sum_{m=1}^{M}\bigl(\hat{y}_m(\vx)_i-\mu_*(\vx)_i\bigr)^{2}.
\label{eq:ens_var}
\end{align}
 Where the training data constrain the prediction tightly, the members agree, and $\sigma_e$ is small, while in regions of input space that the training set represents poorly, the members disagree and $\sigma_e$ grows, providing the signal used to direct further simulation. Each member is a feedforward network with ReLU hidden layers and a linear output head, trained on the mean-squared-error loss with the Adam optimiser~\cite{kingma2014adam}. Use of least-squares loss implies that the only source of inter-member disagreement is the one the ensemble is designed to estimate. The ensemble size $M$ was fixed at $M=12$, chosen through a sensitivity analysis of the stability of the epistemic-uncertainty estimate with respect to $M$, see Fig.~~\ref{fig:ensemble_size}. Based on this result, we concluded that increasing the ensemble size beyond $M=12$ did not return any improved performance.

\begin{figure}[t]
\centering
\includegraphics[width=\textwidth]{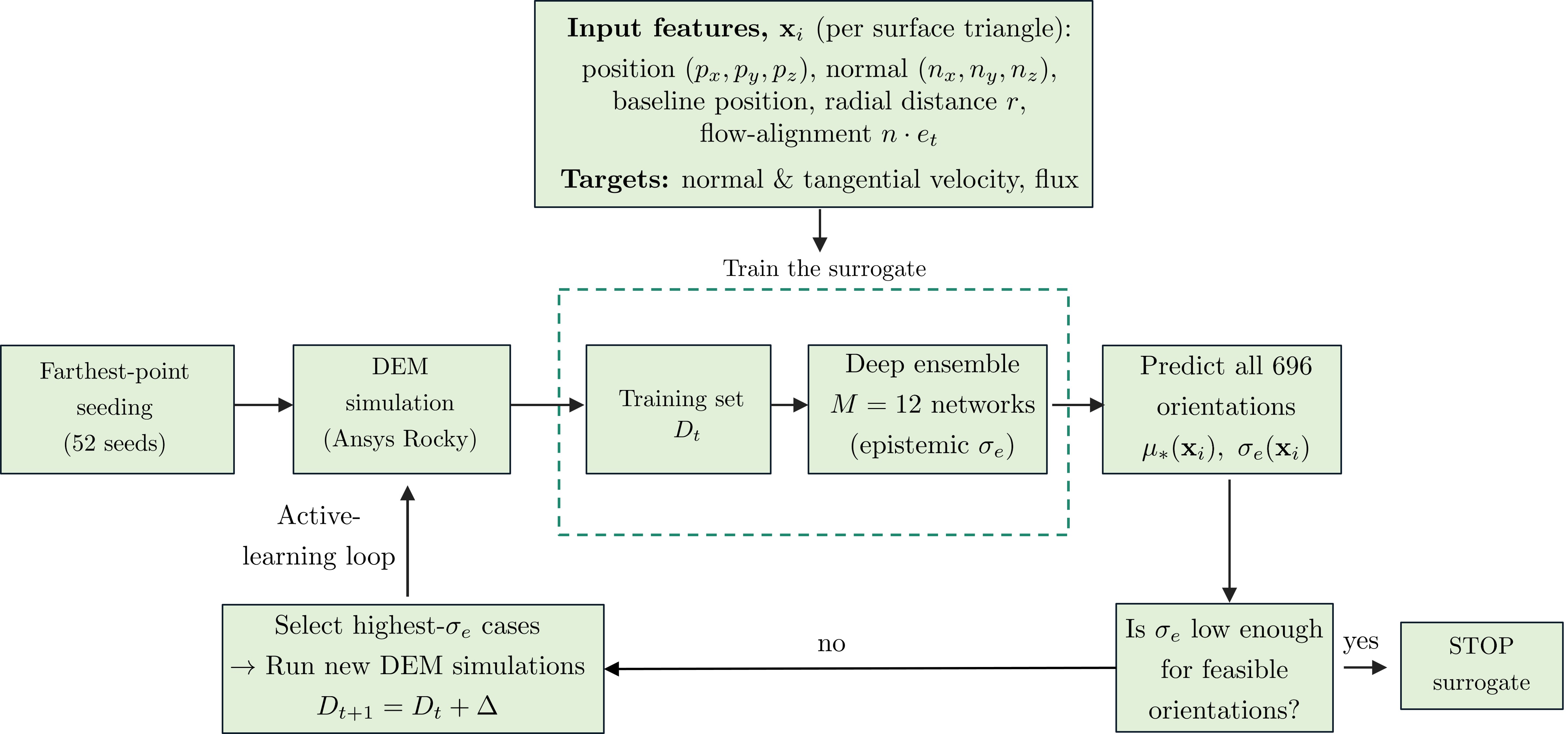}
\caption{Overview of the active-learning surrogate framework. Farthest-point seeding selects the initial orientations, which are simulated with DEM and used to train the deep ensemble. The ensemble predicts the primary fields and an epistemic uncertainty for every feasible orientation. The most uncertain orientations are simulated and added to the training set, and the loop repeats until the uncertainty over the feasible set is acceptably low.}
\label{fig:methodology}
\end{figure}

 The overall workflow of the method, from seeding through active learning ~\cite{ref_al_survey}, is summarised in Fig.~\ref{fig:methodology}.

\subsubsection{Physics-informed input features}\label{sec:features}
Each triangle of each orientation is described by an $11$-dimensional feature vector,
\begin{equation}
\vx_i = [\, p_{i,x}, p_{i,y}, p_{i,z},\; n_{i,x}, n_{i,y}, n_{i,z},\; p^0_{i,x}, p^0_{i,y}, p^0_{i,z},\; r_i,\; \vn_i\!\cdot\!\hat{\mathbf{e}}_t \,]^{\top},
\label{eq:features}
\end{equation}
each component of which is chosen because it controls a specific aspect of the expected collision statistics. The rotated centroid position $\mathbf{p}_i$ locates the facet within the container, which determines the depth in the media bed and the local flow environment the facet is exposed to. The rotated unit normal $\vn_i$ determines the local orientation of a collision, as for a given incoming stream of particles, it fixes the impact angle, which enters the Finnie relation through $f(\gamma)$ and separates grazing impacts from near-normal ones. The baseline coordinates $\mathbf{p}^0_i$ of the triangle in the unrotated configuration act as an orientation-invariant identity of the facet, allowing the network to associate the responses of the same physical facet across different orientations and thereby to generalise a facet's behaviour from the orientations in which it has been observed to those in which it has not. For instance, a facet may occupy a geometrically favourable position for high collision rates, yet still receive fewer impacts because of particle screening near the centre of the geometry. 

The radial distance $r_i$ from the container axis controls the local speed of the bulk media, since the flow is a rotation and the azimuthal velocity of the stream grows with radius. Although $r_i$ is a deterministic function of the position components already present in the feature vector, supplying it directly relieves the network of having to learn this nonlinear combination from the raw coordinates, while also encoding rotational symmetry of the process into the input~\cite{daw2020physics}. This in turn is expected to improve data efficiency when the training set is small. 

Finally, the alignment $\vn_i\!\cdot\!\hat{\mathbf{e}}_t$ between the facet normal and the local azimuthal flow direction carries the windward--leeward distinction, which is the dominant physical factor determining the wear pattern. The sign of this dot product states which way the facet faces relative to the oncoming media. A negative value, $\vn_i\!\cdot\!\hat{\mathbf{e}}_t < 0$ (windward) means the outward normal points against the flow direction, so the facet faces into the oncoming stream and is bombarded continuously. On the other hand,  $\vn_i\!\cdot\!\hat{\mathbf{e}}_t > 0$ (leeward) means the facet points away from the flow so that it is in the shadow of the workpiece, receiving almost no impacts. A value near zero corresponds to a facet nearly parallel to the flow, which is consequently grazed by most particles (low angle of incidence). 

All eleven features are computable from the STL geometry of an orientation alone, without any DEM data, which is what allows the surrogate to be evaluated on orientations that have never been simulated. Additionally, given the generality of the use of STL geometry enables the same procedure to be used for arbitrary complex shaped contours with little change. The chosen features determine the expected per-facet collision statistics, and they do not resolve the stochastic realization of individual collisions in a dense many-body abrasive flow. This sets an irreducible limit on the achievable accuracy (see Sec.~\ref{sec:unc_truth}).

\subsubsection{Data extraction and training targets}\label{sec:data}
For every simulated orientation, the collision data (targets) $v_n$, $v_t$, and $\phi$ for every triangle in the steady-state window are obtained. One simulated orientation therefore contributes $1180$ training rows. For each triangle, the feature vector of Eq.~\ref{eq:features} is paired with its targets. Separate ensembles of identical architecture are trained for each target field. The eleven components of Eq.~\ref{eq:features} are of different numerical scales, such as positions and the radial distance, which are of length units of order millimeters, while the normal components and the flow alignment are dimensionless quantities bounded to $[-1,1]$. Left unscaled, gradient-based training would be dominated by the largest-magnitude features regardless of their physical relevance. Each feature component is therefore standardised to zero mean and unit variance,
\begin{equation}
\tilde{x}_{i,j} = \frac{x_{i,j}-\mu_j}{\sigma_j},
\label{eq:standardise}
\end{equation}
where $\mu_j$ and $\sigma_j$ are the mean and standard deviation of component $j$ taken over all triangles of the training orientations only. These fixed statistics are then applied unchanged to the validation and test-oriented input features, leaving the feasible set data aims to restrict data leakage from the feasible set to the model.

\subsubsection{In-distribution requirement and seed selection}\label{sec:seeds}
A data-driven surrogate is reliable only for inputs that resemble its training data, so before the surrogate is used we require that every feasible orientation be an interpolation of the training set rather than an extrapolation. To evaluate this, we would need to compare complete orientations with each other, whereas Eq.~\ref{eq:features} applies to individual triangles. Each orientation is therefore summarised by a single descriptor $\bm{\psi}(\vx)$, formed from the mean and standard deviation of each feature component taken over all $1180$ triangles of that orientation. This reduces the per-triangle description of an orientation into one fixed-length metric of how the workpiece as a whole is presented to the flow. The components of $\bm{\psi}$ are themselves on different scales, so each is standardised across the feasible set $\mathcal{F}$ in the same manner as Eq.~\ref{eq:standardise}, this time with the mean and standard deviation taken over the $696$ feasible orientations rather than over triangles. Euclidean distances between orientations are then computed in this standardised descriptor space.

The applicability threshold is set from the intrinsic density of the feasible set itself, in the manner of $k$-nearest-neighbour applicability-domain analysis~\cite{sahigara2013defining, sun2022out}. Taking $k=1$, the threshold $d^{\star}$ is the $95$th percentile of nearest-neighbour distances within $\mathcal{F}$, corresponding to the conventional significance level $\alpha=0.05$ of statistical anomaly detection, and for the present geometry gives the dimensionless value $d^{\star}=1.21$. The seed set is then built by farthest-point sampling~\cite{gonzalez1985clustering, sener2017active}, which repeatedly adds the orientation farthest from those already chosen. Seeds are added until the covering radius of the feasible set falls below $d^{\star}$, which occurs at $N_{\mathrm{seed}}=52$ orientations, under $8\%$ of the feasible set. By construction, every feasible orientation then lies within $d^{\star}$ of a simulated orientation. We emphasise that this seeding procedure ensures, \emph{a priori} and from geometry alone, does not result in any configurations requiring extrapolation.

\subsubsection{Execution of the active-learning loop}\label{sec:al_execution}
The loop is initialised with the $52$ farthest-point seed orientations of Section~\ref{sec:seeds} and a fixed held-out validation set of $14$ orientations spanning the feasible space, which are not used for training at any iteration. At iteration $t$ the deep ensemble is trained on the current pool $\mathcal{D}_t$ (with $|\mathcal{D}_1|=52$), predicts the primary fields and the orientation-level uncertainty $\sigma_e$ for all remaining feasible orientations. The $10$ orientations with the largest $\sigma_e$ (see, Appendix \ref{appendix:sigma_ori}) are then simulated with DEM and added to the training pool, giving $|\mathcal{D}_{t+1}|=|\mathcal{D}_t|+10$, and the ensemble is retrained. Five iterations are run in total, ending with a surrogate trained on $92$ DEM simulations, about $13\%$ of the feasible set (Table~\ref{tab:al_schedule}). All results in Section~\ref{sec:results} are indexed by this schedule: iteration~$1$ is the surrogate trained on the seeds alone, and iteration~$5$ is the fully refined surrogate.

\begin{table}[ht]
\centering
\caption{Schedule of the active-learning loop. At each iteration the ensemble is trained on the current pool, applied to the held-out validation set and the remaining feasible orientations, and the $10$ highest-uncertainty orientations are simulated and added to the pool. The last two columns report the range of orientation-level epistemic uncertainty $\sigma_e$ across the remaining feasible orientations, for the normal ($v_n$) and tangential ($v_t$) impact velocity components.}
\label{tab:al_schedule}
\setlength{\tabcolsep}{4pt}
\begin{tabular}{c@{\hskip 6pt}c@{\hskip 6pt}c@{\hskip 6pt}c@{\hskip 10pt}c@{\hskip 10pt}c}
\toprule
Iteration & $|\mathcal{D}_t|$ & Validation & Remaining & $\sigma_{e,v_n}\times 10^{-3}$ & $\sigma_{e,v_t}\times 10^{3}$ \\
\midrule
$1$ & $52$ & $14$ & $644$ & $2.41$--$36.3$ & $4.16$--$52.2$ \\
$2$ & $62$ & $14$ & $634$ & $2.18$--$14.3$ & $4.33$--$29.3$ \\
$3$ & $72$ & $14$ & $624$ & $1.57$--$9.91$ & $2.80$--$34.7$ \\
$4$ & $82$ & $14$ & $614$ & $1.06$--$7.09$ & $1.72$--$28.1$ \\
$5$ & $92$ & $14$ & $604$ & $1.22$--$7.04$ & $1.94$--$21.8$ \\
\bottomrule
\end{tabular}
\end{table}

\section{Results and Discussion}\label{sec:results}

The active-learning loop of Section~\ref{sec:al_execution} is run on the flat-wall workpiece for five iterations, and the results from the surrogate model are reported in this section. We first establish that the ensemble's epistemic uncertainty is a truthful predictor of the realised error (Section~\ref{sec:unc_truth}). The evaluation metrics and the uncertainty-band classification are introduced there, where they are first used. We then track the impact-velocity surrogate across iterations (Section~\ref{sec:vel_results}), do the same for the particle impact flux (Section~\ref{sec:flux_results}), and close by combining the predicted primary fields through the Finnie relation to compare reconstructed and DEM wear patterns at representative orientations from each uncertainty band (Section~\ref{sec:wear}).

\subsection{Truthfulness of the predicted uncertainty}\label{sec:unc_truth}

Two metrics are computed across the $N$ triangles of each held-out orientation. The coefficient of determination, given by
\begin{equation}
R^2 = 1-\frac{\sum_i (y_i-\hat{y}_i)^2}{\sum_i (y_i-\bar{y})^2},
\label{eq:r2}
\end{equation}
measures how much of the spatial variance of the DEM field is captured by the prediction. The Spearman rank correlation~\cite{spearman1904proof, sai2020fatigue},
\begin{equation}
\rspear = 1-\frac{6\sum_i d_i^2}{N(N^2-1)},
\qquad d_i = \operatorname{rk}(y_i)-\operatorname{rk}(\hat{y}_i),
\label{eq:spearman}
\end{equation}
compares only the orderings of the triangles, where $\operatorname{rk}(\cdot)$ is the rank of a triangle when all $N$ are sorted by the corresponding field. Both are reported because neither alone is sufficient. For this problem $\rspear$ is the critical metric. We note that the main aim of the surrogate is to reproduce the spatial wear pattern, i.e., which facets erode strongly and which are relatively less-worn. A model with a systematic scale error scores poorly on $R^2$ while reproducing the pattern perfectly. On the other hand, a respectable $R^2$ can be earned from the bright exposed band alone, while the ordering of the many interior facets is randomized. The metric $\rspear$ is hence insensitive to monotone rescaling, but can capture reordering failures, while being bounded to $[-1,1]$. In stark contrast, $R^2$ is unbounded below and a single badly-predicted orientation can dominate any average value.

\begin{figure}[H]
\centering
\begin{minipage}{0.48\textwidth}
  \centering
  \includegraphics[width=\linewidth]{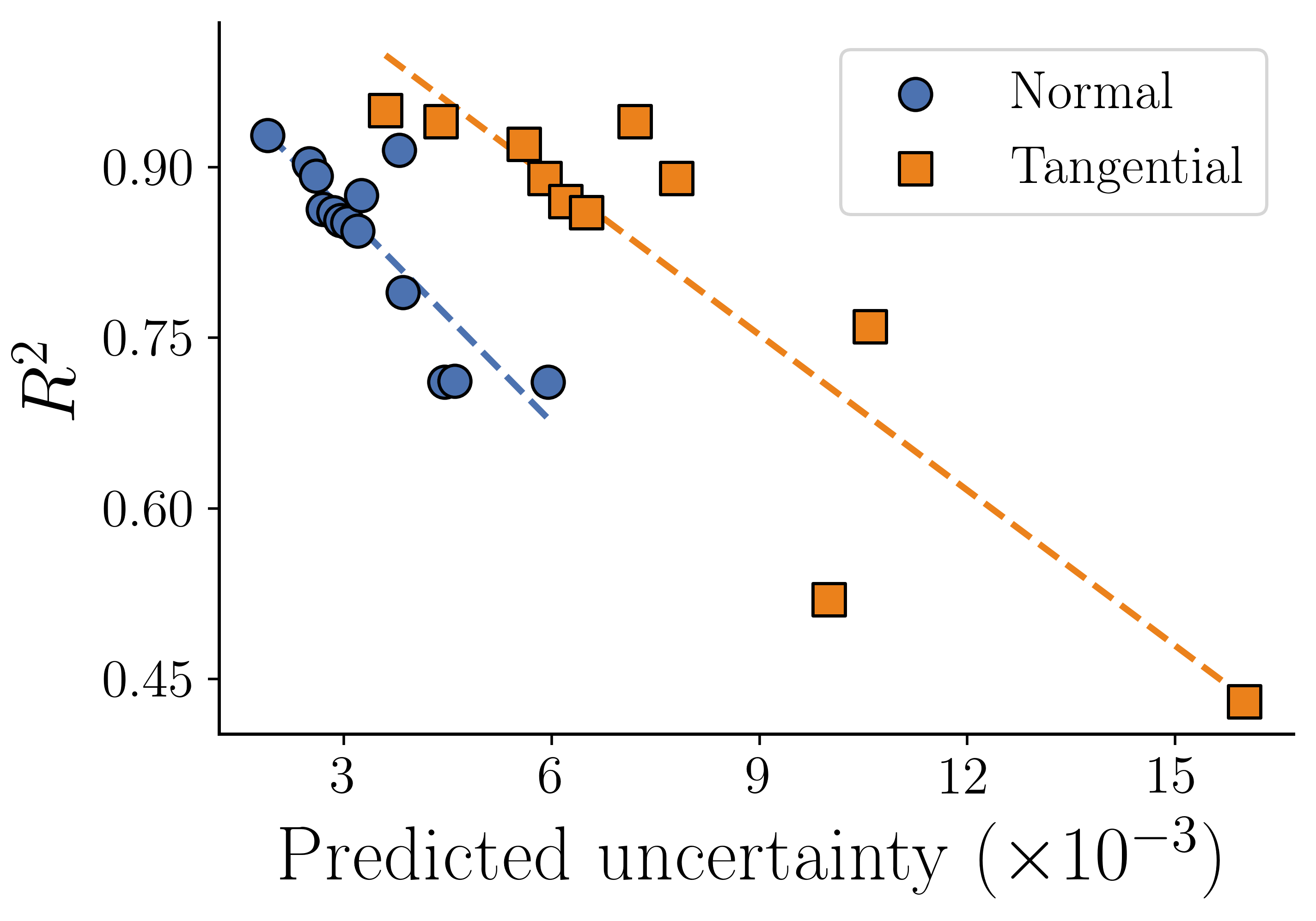}\\
  \footnotesize (a)
\end{minipage}\hfill
\begin{minipage}{0.48\textwidth}
  \centering
  \includegraphics[width=\linewidth]{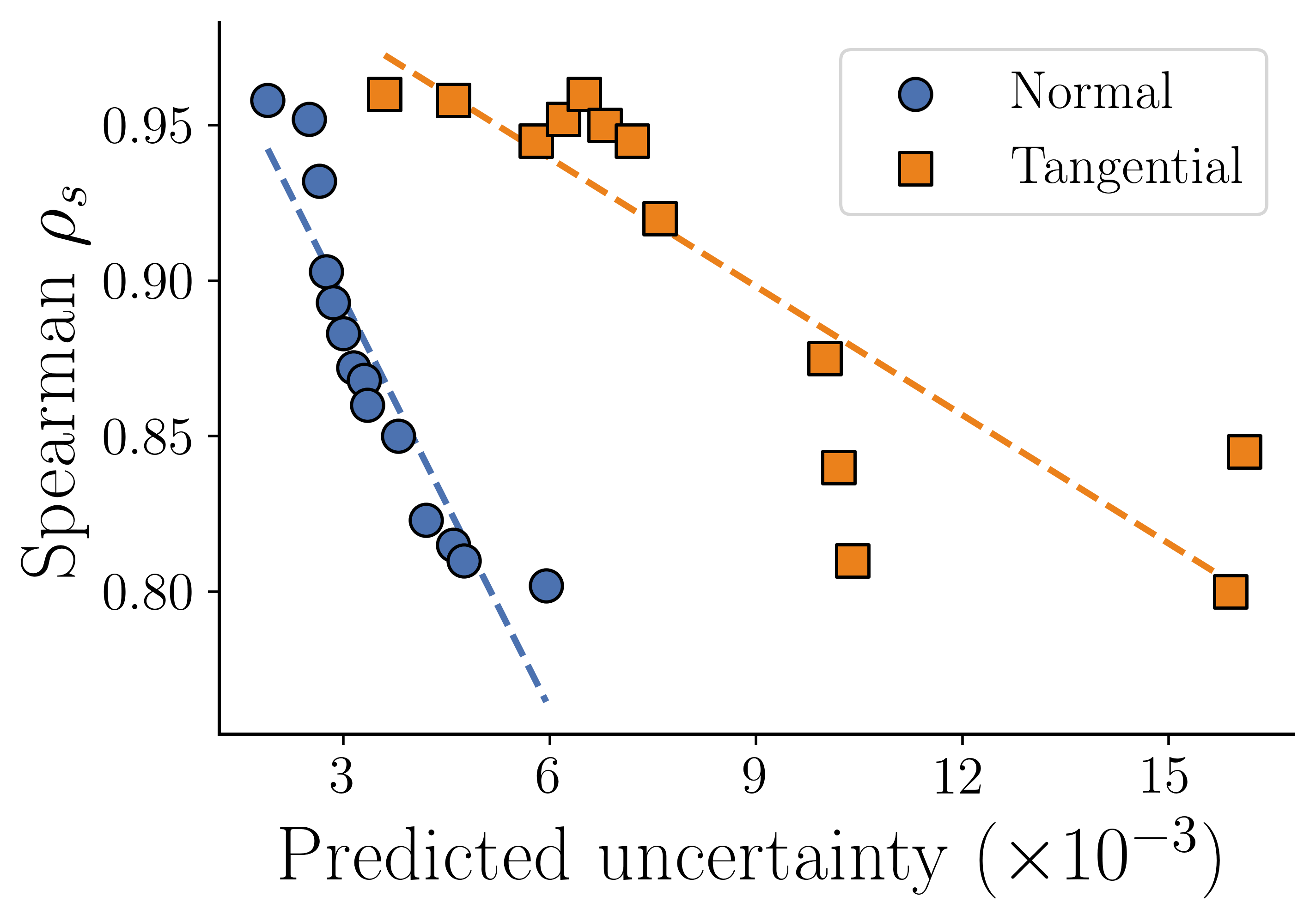}\\
  \footnotesize (b)
\end{minipage}
\caption{Truthfulness of the predicted uncertainty on the held-out orientations of the final iteration. Each point is one orientation; the surface-averaged epistemic uncertainty $\sigma_e(\vx)$ is plotted against (a) the achieved $R^2$ and (b) the achieved $\rspear$ for the normal and tangential velocity components. Accuracy falls as predicted uncertainty rises, with rank correlations of $-0.77$ ($p=0.001$) and $-0.82$ ($p=0.002$) for the normal and tangential components respectively, establishing the uncertainty as a truthful, pre-DEM predictor of where the surrogate is least reliable.}
\label{fig:truthfulness}
\end{figure}

The premise of the loop is that the ensemble's inter-member disagreement identifies where the surrogate is likely to be wrong. Before this signal is used to determine additional simulations, uncertainty of the predictions must be shown to be truthful. This implies that the predicted uncertainty and realized error must move together across held out orientations. A test of this on the 14 held-out orientations of the final iteration is presented in Fig.~\ref{fig:truthfulness}. This figure plots the surface-averaged (over all triangles for a given orientation) epistemic uncertainty $\sigma_e(\vx)$ (see Eq.~\ref{eq:sigma_ori}) against the achieved $R^2$ and $\rspear$. Both panels show the expected trend as uncertainty rises, accuracy decreases monotonically in the ranking sense. Spearman correlations between $\sigma_e$ and $R^2$ are $-0.77$ ($p=0.001$) for the normal component and $-0.82$ ($p=0.002$) for the tangential component. Orientations the model predicts as being uncertain are, once DEM is run, the orientations for which it provides the worst prediction. Since $\sigma_e$ is computed before any DEM result exists for the orientation in question, and is available for all $696$ feasible orientations, this quantity reliably determines the configuration for which a subsequent DEM simulation would be most informative.

The truthfulness trend of Fig.~\ref{fig:truthfulness} is also what allows every feasible orientation to be classified by its expected accuracy. Three bands are defined at fixed accuracy levels,
\begin{equation}
\text{Low: } \rspear \ge 0.90,
\qquad
\text{Medium: } 0.80 \le \rspear < 0.90,
\qquad
\text{High: } \rspear < 0.80.
\label{eq:bands_rho}
\end{equation}
At $\rspear\ge0.90$ the predicted and DEM orderings agree almost everywhere and the identity of the most- and least-eroded regions is nearly the same. Between $0.80$ and $0.90$ the gross exposed--sheltered structure survives while there is significant deviation in finer ordering within regions. Below $0.80$ the ordering is no longer reliable for identifying individual high-wear facets, and a further DEM simulation is warranted for this orientation.

These bands are defined in $\rspear$, which requires DEM ground truth and so cannot be evaluated for the feasible orientations that are never simulated. The epistemic uncertainty $\sigma_e$, by contrast, is available for all of them, so the bands must be expressed as thresholds on $\sigma_e$. To do this we fit the relationship between $\sigma_e$ and the attained $\rspear$ on the held-out orientations and read off the $\sigma_e$ values at which the fitted curve crosses $0.90$ and $0.80$. The fit is performed on the final iteration, iteration~$5$, since it is the most accurate model and therefore yields the most reliable $\sigma_e$--$\rspear$ relationship. Constraining the thresholds to this single fit, in place of continuous re-fitting at each iteration, ensures that a given band denotes the same predictive accuracy throughout. The growth of the low-$\sigma_e$ population across iterations therefore reflects an actual improvement of the model against a fixed accuracy standard, rather than a shift in an iteration-dependent definition. This gives thresholds of $(4,7)\times10^{-3}$ for the normal component and $(10,15)\times10^{-3}$ for the tangential component. These two $\sigma_e$ values are then fixed and applied unchanged to every iteration. For each field, they are the thresholds used to classify all remaining feasible orientations. What is held constant is therefore the accuracy level of Eq.~\ref{eq:bands_rho}, and not the $\sigma_e$ value at which it occurs. Consequently, the acquisition step uses only the raw $\sigma_e$ ranking and requires no thresholds.

\subsection{Impact-velocity prediction}\label{sec:vel_results}

\begin{figure}[H]
\centering
\begin{minipage}{0.48\textwidth}
  \centering
  \includegraphics[width=\linewidth]{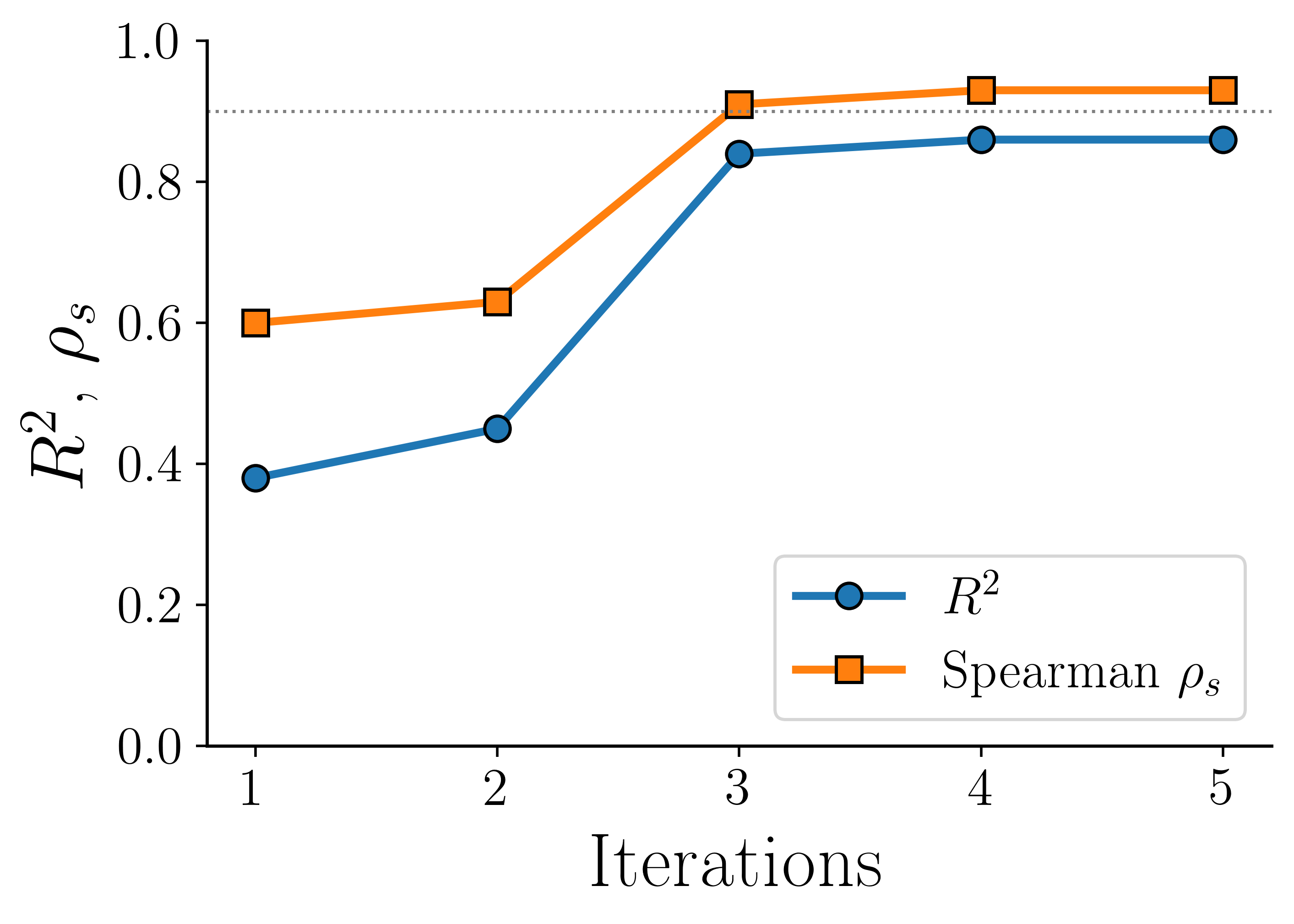}\\
  \footnotesize (a) Normal velocity, held-out validation
\end{minipage}\hfill
\begin{minipage}{0.48\textwidth}
  \centering
  \includegraphics[width=\linewidth]{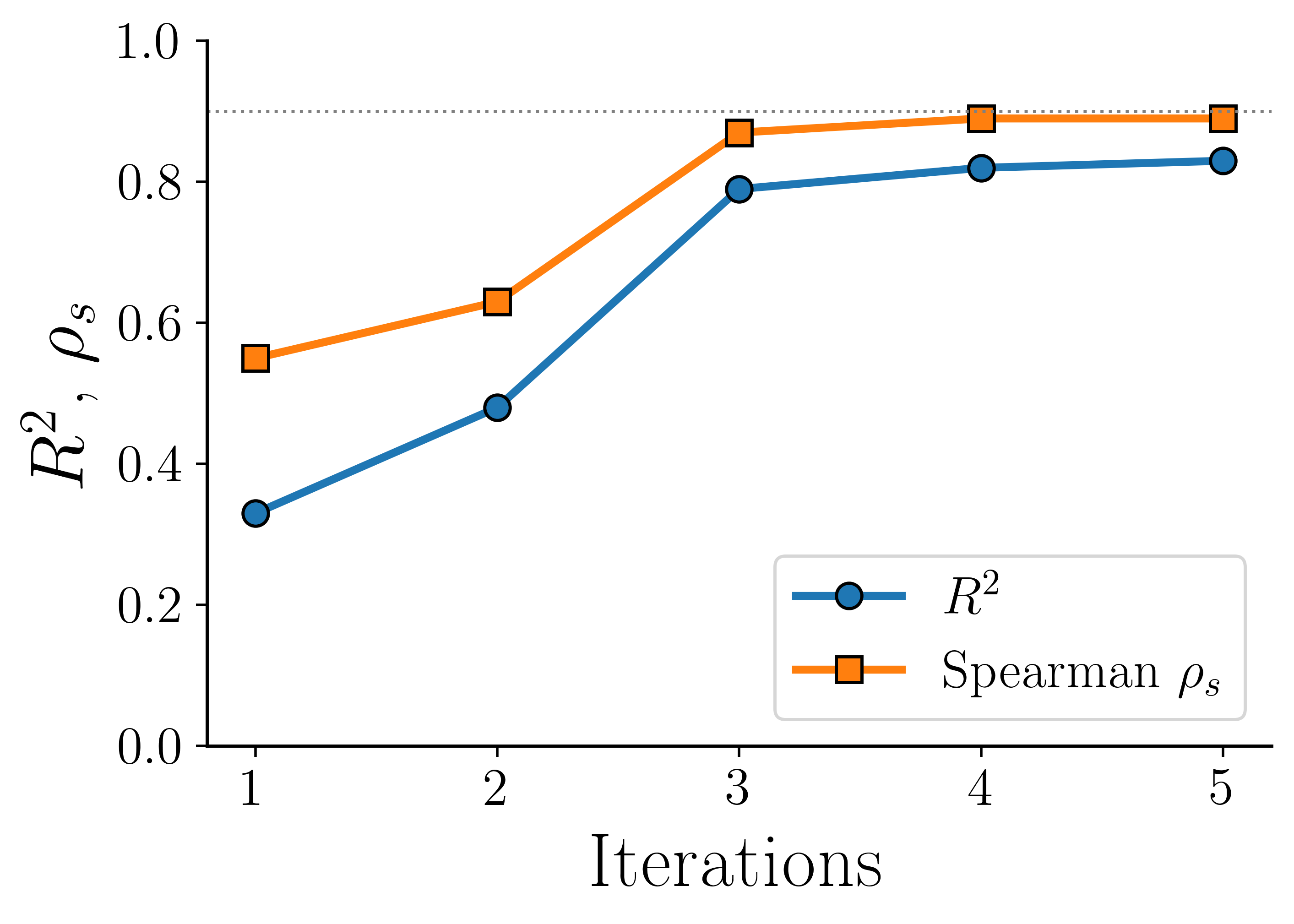}\\
  \footnotesize (b) Tangential velocity, held-out validation 
\end{minipage}
\vspace{6pt}

\begin{minipage}{0.48\textwidth}
  \centering
  \includegraphics[width=\linewidth]{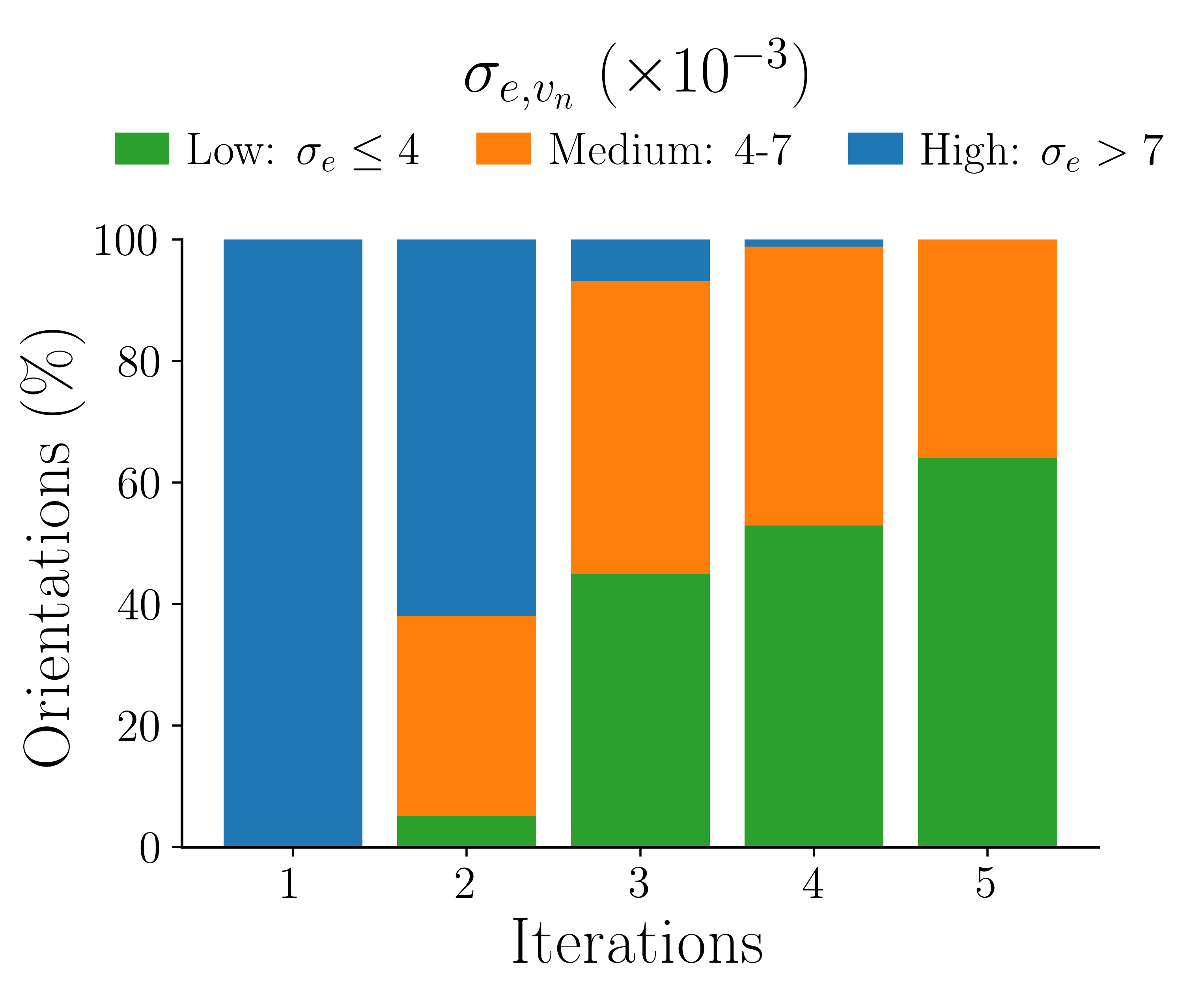}\\
  \footnotesize (c) Normal velocity, uncertainty bands 
\end{minipage}\hfill
\begin{minipage}{0.48\textwidth}
  \centering
  \includegraphics[width=\linewidth]{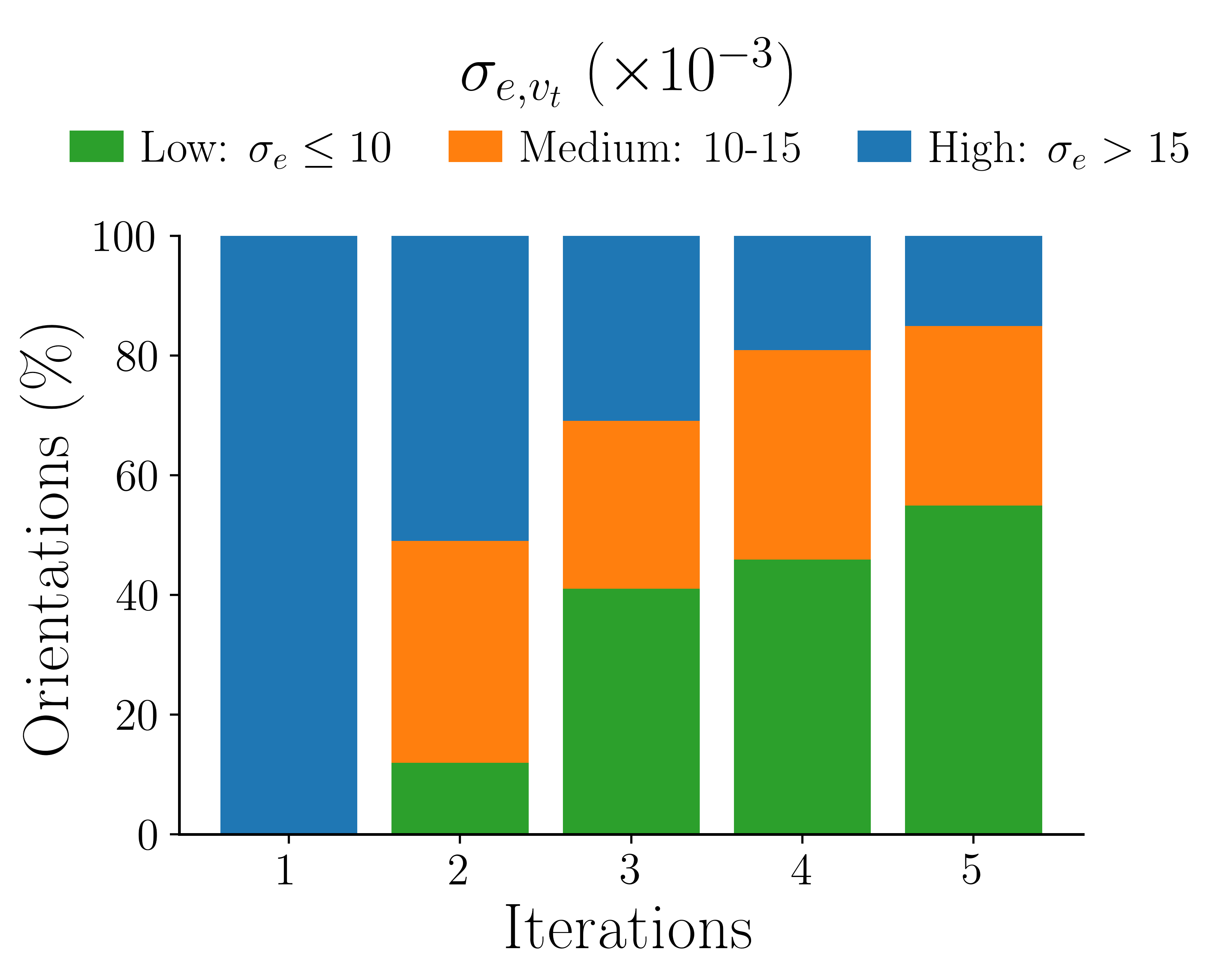}\\
  \footnotesize (d) Tangential velocity, uncertainty bands
\end{minipage}
\caption{Impact-velocity surrogate across the five active-learning iterations, for the normal (left column) and tangential (right column) components. Panels~(a) and (b) show the orientation-averaged $R^2$ and $\rspear$ on the $14$ held-out validation orientations; the dashed grey line marks $R^2=0.9$. Held-out accuracy saturates after iteration~$3$ while the low-uncertainty band continues to grow through iteration~$5$, reflecting that later iterations extend confident coverage across the feasible space rather than improving accuracy on the already-covered held-out set. Panels~(c) and (d) classify all remaining feasible orientations into the low-, medium-, and high-uncertainty bands of Eq.~\ref{eq:bands_rho}, using the fixed field-specific $\sigma_e$ thresholds of Section~\ref{sec:unc_truth}. }
\label{fig:impact_velocity}
\end{figure}

The model predicts, per surface triangle, the mean of the normal and tangential impact-velocity components. Panels~(a) and (b) of Fig.~\ref{fig:impact_velocity} track the orientation-averaged (over N triangles) $R^2$ and $\rspear$ on the $14$ held-out orientations across the five iterations. The per-triangle normal impact velocities predicted by the model, together with the corresponding DEM results and prediction uncertainty for a representative held-out orientation from the third iteration, are presented in Appendix~\ref{appendix:per_tri}. For the normal component $R^2$ rises from $0.38$ at iteration~$1$ to $0.86$ at iteration~$5$ and $\rspear$ from $0.60$ to $0.93$; for the tangential component $R^2$ rises from $0.33$ to $0.83$ and $\rspear$ from $0.55$ to $0.89$. The large gain at iteration~$3$ occurs because the orientations added there filled a sparsely-sampled region of feasible space in whose neighbourhood several held-out orientations already exist. Once training data appear near them, predictions improve significantly. After iteration~$3$ the curves flatten, changing by no more than $0.02$ between successive iterations. As mentioned before, the chosen seed density prevents extrapolation, while active learning increases density where the fields vary fastest.

Panels~(c) and (d) of Fig.~\ref{fig:impact_velocity} classify all remaining feasible orientations into the bands of Eq.~\ref{eq:bands_rho} using the field-specific $\sigma_e$ thresholds introduced in Sec.~\ref{sec:unc_truth}. These are again maintained constant across iterations so that the bars are directly comparable. For the normal component, almost no orientations occur in the low band at iteration~$1$ and the high band holds nearly the entire feasible set. By iteration~$3$ the low band reaches about $45\%$ with the high band essentially eliminated. At iteration~$5$ it holds roughly $64\%$, with the remainder sitting in the medium band. The tangential component starts more uncertain and contracts more slowly, its low band growing from below $5\%$ through $13\%$, $41\%$, and $46\%$ to about $55\%$, while its high band shrinks to about $15\%$. In both components the low band grows monotonically at every iteration and the high band contracts monotonically, so each batch of added DEM simulations moves a further share of the feasible space into the confident region. The tangential component lags the normal one throughout, consistent with its lower predictive accuracy.

The loop is stopped at iteration~$5$ for practical reasons, despite the fact that a substantial fraction of orientations still sits in the medium band for the normal component, and the tangential component retains a high-uncertainty region. The trend across the five iterations is nonetheless clear and uninterrupted. Each iteration adds only ten DEM simulations, yet produces a visible increase in the low-band population. Neither component shows any sign of this growth saturating by iteration~$5$. Extrapolating this behaviour, a modest number of further iterations would be expected to bring the remaining medium- and high-uncertainty orientations into the low band for both components. The broader implication of this result is that the surrogate can be driven to confident, accurate predictions over the entire feasible set at a small fraction of the cost of exhaustive simulation, with accuracy improving steadily as each small batch of simulations is added.

\subsection{Particle-flux prediction}\label{sec:flux_results}

\begin{figure}[H]
\centering
\begin{minipage}{0.48\textwidth}
  \centering
  \includegraphics[width=\linewidth]{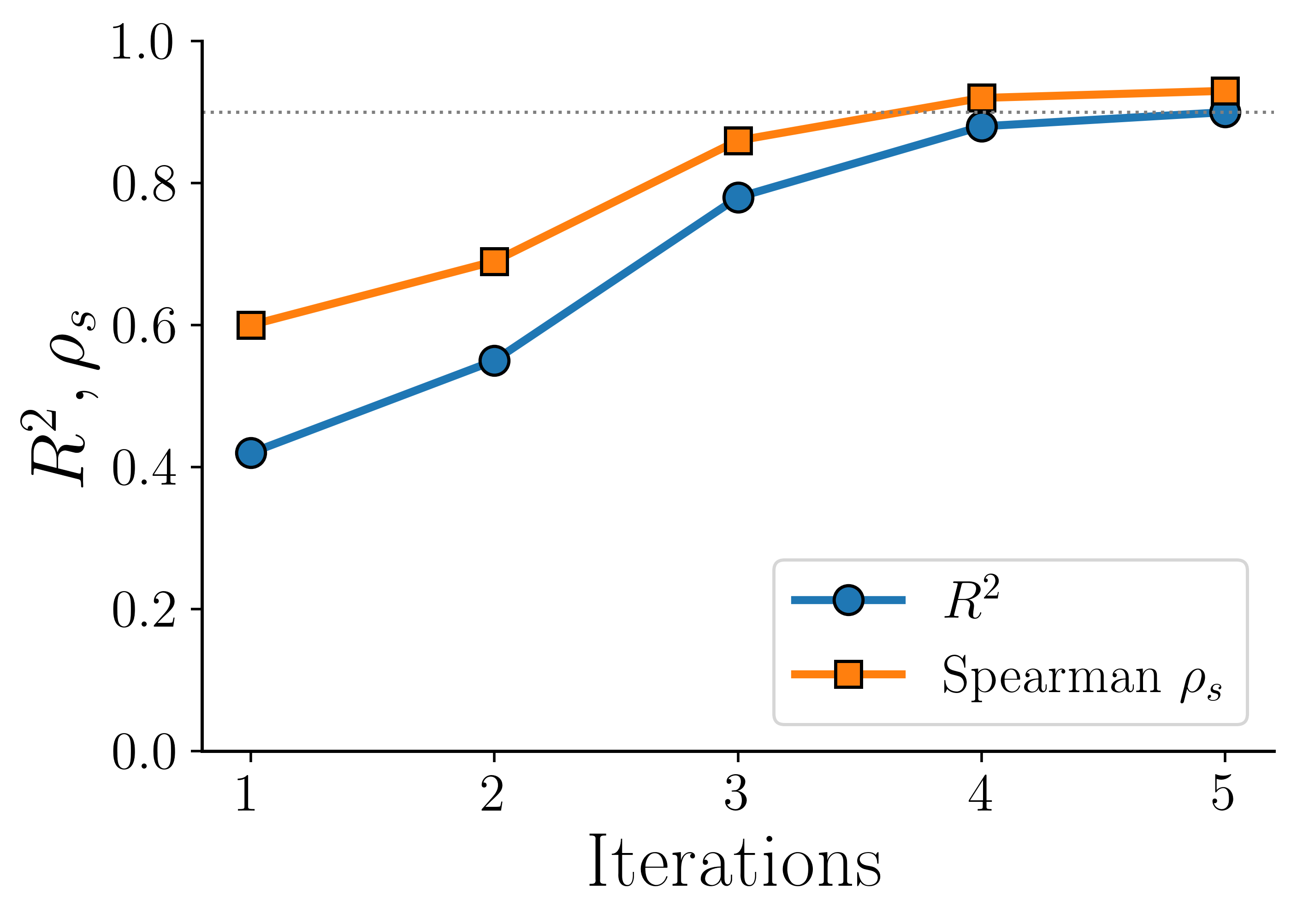}\\
  \footnotesize (a) Held-out validation
\end{minipage}\hfill
\begin{minipage}{0.48\textwidth}
  \centering
  \includegraphics[width=\linewidth]{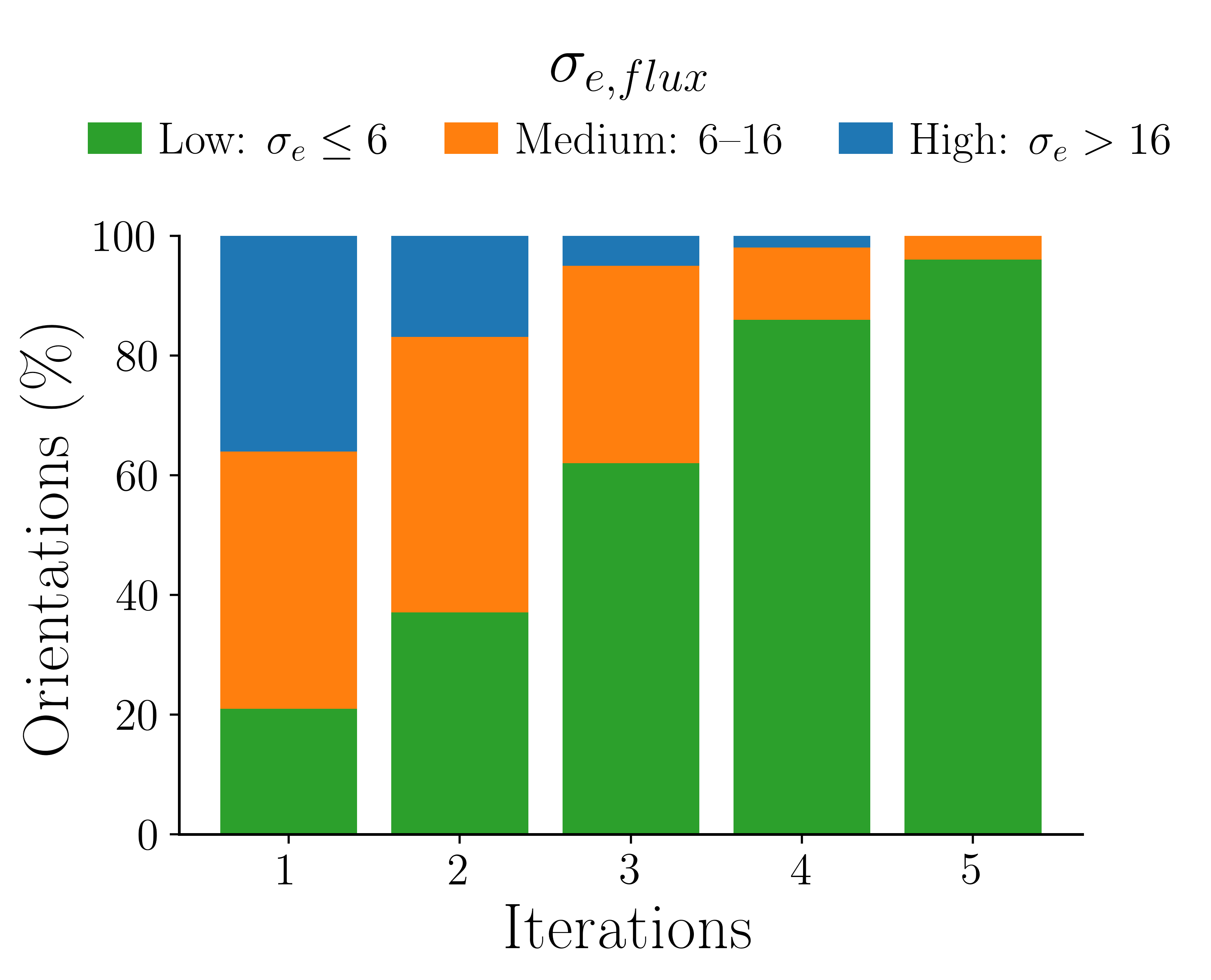}\\
  \footnotesize (b) Uncertainty bands
\end{minipage}
\caption{Particle-flux surrogate across the five active-learning iterations. (a) Orientation-averaged $R^2$ and $\rspear$ on the held-out validation orientations, rising from $0.42$ and $0.60$ at iteration~$1$ to $0.90$ and $0.93$ at iteration~$5$. (b) Classification of all remaining feasible orientations into the uncertainty bands of Eq.~\ref{eq:bands_rho}; the low-uncertainty population grows as targeted DEM data are added.}
\label{fig:flux}
\end{figure}

The particle impact flux is the third primary field of the Finnie relation, and its prediction follows the same trend as the velocities described in the previous section, see Fig.~\ref{fig:flux}. Panel (a) in this figure shows the held-out accuracy across iterations. The average $R^2$ rises from $0.42$ at iteration~$1$ to $0.90$ at iteration~$5$, and the average $\rspear$ from $0.60$ to $0.93$, with the same sharp increase at iteration~$3$ and subsequent saturation observed for the velocities. The flux is predicted somewhat more accurately than the tangential velocity at the fifth iteration, because the particle flux depends more smoothly on orientation when compared to the velocity components. 
Fig.~\ref{fig:flux}(b) classifies the feasible set by flux uncertainty using the same $\rspear$-based construction of Eq.~\ref{eq:bands_rho}. It is clear that the low band grows monotonically across iterations while the high band decreases, following the velocity components. All three primary fields therefore reach the same state by iteration~$5$.
\subsection{Wear-field reconstruction through the Finnie relation}\label{sec:wear}

\begin{figure}[t]
\centering
\includegraphics[width=\textwidth]{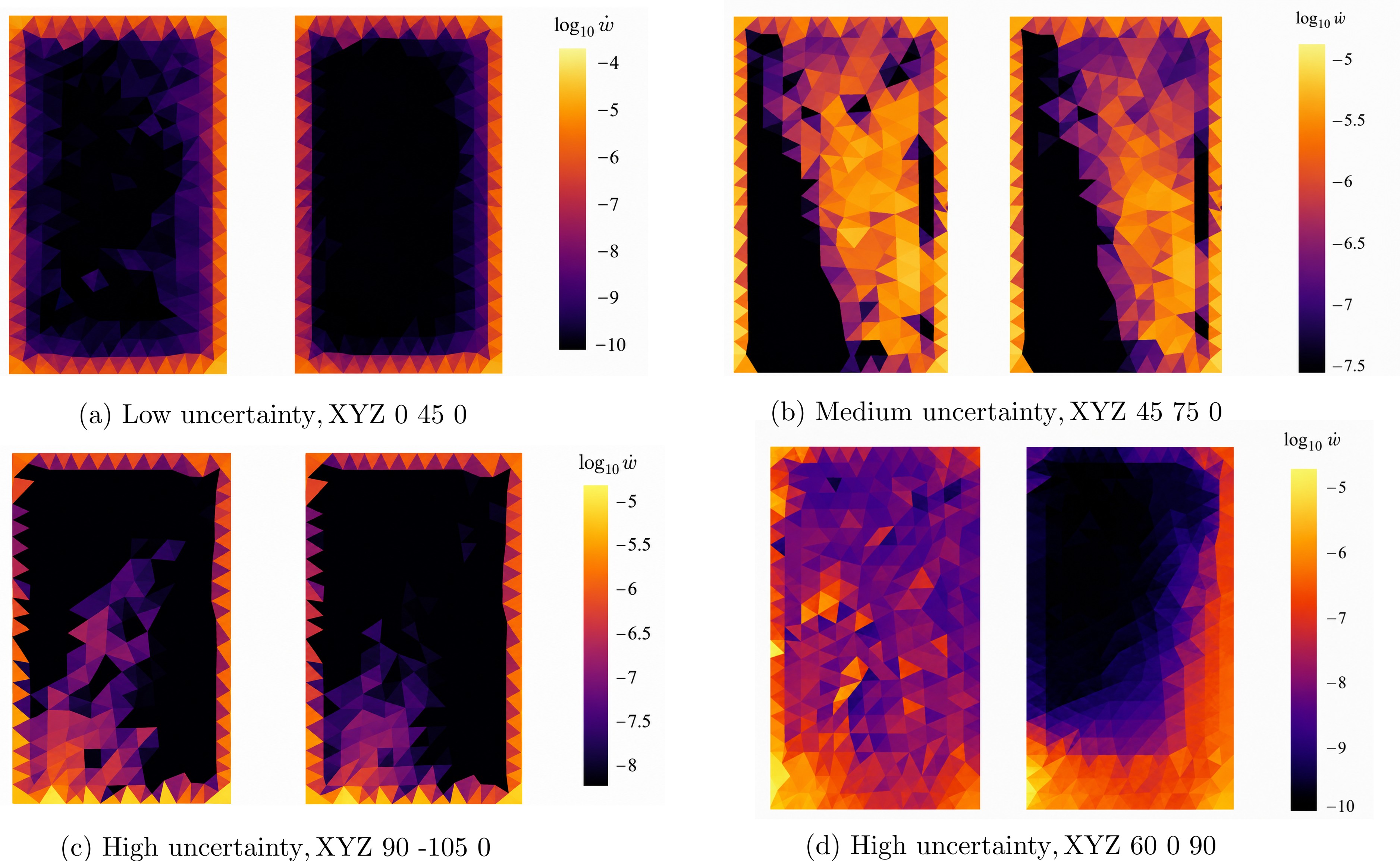}
\caption{Finnie-reconstructed wear fields compared with DEM for four representative orientations spanning the uncertainty bands: (a) low uncertainty, \text{XYZ 0 45 0}; (b) medium uncertainty, \text{XYZ 45 75 0}; (c) high uncertainty, \text{XYZ 90 -105 0}; and (d) high uncertainty, \text{XYZ 60 0 90}. Within each panel the left map is the DEM wear rate and the right map is the model wear rate reconstructed by feeding the predicted primary fields through the Finnie relation, with a shared $\log_{10}\dot{w}$ colour scale per panel. Agreement is close in the low band and degrades progressively through the medium and high bands, showing that the epistemic uncertainty of the primary fields predicts the fidelity of the reconstructed wear pattern.}
\label{fig:wear_maps}
\end{figure}

Combining the predicted velocity and flux fields through the Finnie relation (Eq.~\ref{eq:finnie_model}) yields the wear field without regressing wear directly, confining the sharp angular non-linearity and squared-velocity dependence to the analytical equation (Section~\ref{sec:why_primary}). The wear field is reconstructed from the predicted primary fields, so its fidelity should follow their epistemic uncertainty. To test this, four representative orientations are selected spanning the uncertainty bands, requiring in each case that the orientation sits in its band consistently across all primary fields. Fig.~\ref{fig:wear_maps} shows the corresponding DEM-predicted wear map beside the reconstructed map for each. The panels in a row  share a common $\log_{10}$ color scale, so agreement can be understood directly from the similarity of the patterns.

For the low-uncertainty orientation, the reconstruction is nearly indistinguishable from DEM: the high-wear band around the exposed edges, the sheltered interior, and the most eroded corners are all reproduced, with $R^2=0.90$ and $\rspear=0.97$. For the medium-uncertainty orientation, the gross exposed--sheltered structure survives while the finer detail within transition regions is smoothed; the rank agreement nonetheless remains high ($\rspear=0.87$). For the two high-uncertainty orientations, the agreement breaks down progressively. The brightest edge facets are still placed roughly correctly, which is why $\rspear$ does not collapse entirely, but the internal structure is largely lost and the magnitudes are no longer meaningful, with $R^2$ falling to $0.11$ for the most uncertain case. Table~\ref{tab:wear_repr} reports, for the four representative orientations, the primary-field uncertainties alongside the wear agreement; from low band to the most uncertain orientation, $R^2$ reduces from $0.90$ to $0.11$, $\rspear$ from $0.97$ to $0.35$, and the hotspot overlap from $0.92$ to $0.46$.
\begin{table}[t]
\centering
\caption{Epistemic uncertainty of the primary fields and agreement of the Finnie-reconstructed wear field with DEM for the four representative orientations of Fig.~\ref{fig:wear_maps}. $\sigma_e$ values are surface-averaged epistemic standard deviations ($\times 10^{-3}$).}
\label{tab:wear_repr}
\renewcommand{\arraystretch}{1.25}
\setlength{\tabcolsep}{6pt}
\begin{tabular}{@{}ll ccc @{\hspace{1.5em}} >{\centering\arraybackslash}p{3.6cm} >{\centering\arraybackslash}p{1.6cm}@{}}
\toprule
& & \multicolumn{3}{c}{\textbf{Primary-field uncertainty}}
  & \multicolumn{2}{c}{\textbf{Reconstructed wear vs DEM}} \\
\cmidrule(lr){3-5} \cmidrule(lr){6-7}
Orientation & Band &
$\sigma_{e,v_n}$ & $\sigma_{e,v_t}$ & $\sigma_{e,flux}$ &
$R^2$ & $\rspear$ \\
\midrule
\text{0\ 45\ 0}    & low    & 1.33 & 2.57  & 4.25  & 0.90 & 0.97 \\
\text{45\ 75\ 0}   & medium & 5.55 & 12.92 & 6.50  & 0.82 & 0.87 \\
\text{90\ -105\ 0} & high   & 5.05 & 17.82 & 8.50  & 0.31 & 0.51 \\
\text{60\ 0\ 90}   & high   & 6.98 & 29.99 & 12.60 & 0.11 & 0.35 \\
\bottomrule
\end{tabular}
\end{table}

Two further observations follow from these results. Firstly, the wear reconstruction degrades faster than the velocity predictions that are used for its construction. This is because an orientation with poor velocity accuracy can yield a much worse wear field, given the quadratic dependence of the latter on the former. This is the quantitative counterpart of the argument in Section~\ref{sec:why_primary} for predicting the smooth fields rather than the wear itself. Secondly, and of greater practical significance, $\rspear$ and the hotspot overlap reduce more weakly than $R^2$. Even at high uncertainty, the model retains partial information about which regions within the workpiece erode, while losing the ability to predict the wear value itself. For a part orientation decision that only compares orientations and their relative wear rates, this residual ranking ability is sufficient. The expected reliability of a reconstructed wear map can therefore be read from the uncertainty of the fields used for its determination.

\section{Summary}\label{sec:conclusion}

This work addressed the cost of mapping orientation-dependent wear in stream finishing, where DEM resolves the wear field of one orientation accurately but cannot be run over the several hundred feasible orientations of a new geometry. The approach rests on three choices. Firstly, the surrogate predicts the smooth primary impact fields, the normal and tangential velocity and the particle flux, and confines the sharp non-linearity of erosion to the analytical Finnie relation, rather than regressing the wear field directly. Secondly, a deep ensemble provides an epistemic uncertainty alongside every prediction. Third, this uncertainty guides an active-learning loop that spends each batch of DEM simulations on the orientations where the model is weakest.

Trained on $92$ simulations, about $13\%$ of the $696$ feasible orientations, the surrogate reached held-out rank correlations of $0.93$ and $0.89$ for the velocity components and $0.93$ for the flux. The central result is that the reported uncertainty is truthful: it anticipates the realised error across held-out orientations ($\rspear=-0.77$ and $-0.82$ for the two velocity components) and thereby predicts, before any DEM is run, how faithful the downstream Finnie wear reconstruction will be, from near-exact agreement ($\rspear=0.97$) at low-uncertainty orientations to a controlled loss of interior detail at high uncertainty. The uncertainty classification places $64\%$ and $55\%$ of the feasible set in the low-uncertainty band for the normal and tangential components after five iterations, with both bands still growing, so a modest number of further iterations is expected to bring the full feasible set to confident coverage.

We emphasize in closing that the framework presented is not specific to the geometry studied here. The input features are computable from a triangulated mesh description of any shape and at any orientation since the  seeding and coverage arguments are purely geometric. Consequently, the uncertainty bands transfer to any field through the same construction. Two natural extensions follow from this work as a result of this generality. The first is to apply the approach to complex industrial geometries, where the feasible orientation set is larger and the wear patterns are more varied. The second is to couple the predicted wear fields to the downstream orientation-sequencing decision itself. We are presenrlty exploring these possibilities and hope to present our results in the near future.
\appendix
\section*{Appendix}

\section{From triangle-level to orientation-level uncertainty}\label{appendix:sigma_ori}
Decisions about which orientation to simulate next are made at the level of the orientation, so the per-triangle uncertainty field is reduced to a single number by the plain arithmetic mean of the per-triangle epistemic uncertainty,
\begin{equation}
\sigma_e(\vx) = \frac{1}{N}\sum_{i=1}^{N}\sigma_{e,i}(\vx_i),
\label{eq:sigma_ori}
\end{equation}
The square root is taken per triangle before the average is formed, where $\sigma_{e,i}(\vx_i)$ is the epistemic uncertainty of a single triangle for a given orientation and $\sigma_{e}(\vx)$ is the average uncertainty over $N$ triangles for a single orientation.

\begin{figure}[H]
\centering
\includegraphics[width=0.6\textwidth]{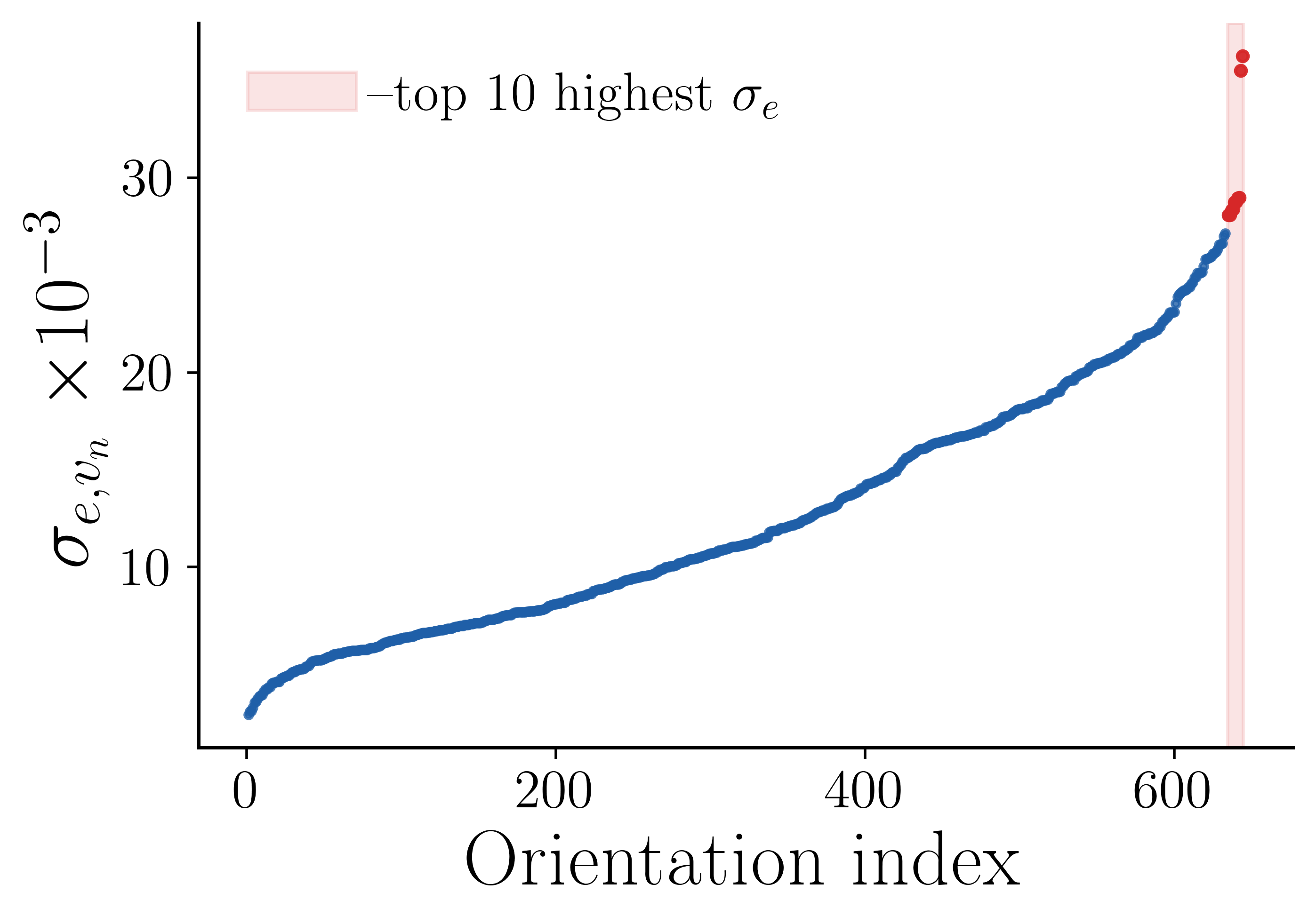}
\caption{Orientation-level epistemic uncertainty $\sigma_{e,v_n}$ (Eq.~\ref{eq:sigma_ori}) for the $644$ feasible orientations not used in training at iteration~$1$, sorted in ascending order. The shaded band marks the highest-uncertainty orientations, which are selected for DEM simulation and added to the training pool for the next iteration.}
\label{fig:unc_all_orientations}
\end{figure}

The average uncertainty measure $\sigma_{e,v_n}$ over $N$ triangles, evaluated using Eq.~\ref{eq:sigma_ori} for the output parameter normal velocity, is shown in Fig.~\ref{fig:unc_all_orientations} for all $644$ feasible orientations not used for training at iteration~1 of the active-learning loop, sorted in ascending order. The uncertainty varies smoothly by more than a factor of two across the feasible set, from about $6\times10^{-3}$ to $14\times10^{-3}$. The surrogate therefore discriminates clearly between orientations it has already learned well and orientations for which its training data are still thin. The shaded band at the right marks the orientations with the largest $\sigma_{e,v_n}$, which are the ones selected for DEM simulation at the next iteration and added to the training pool for iteration~$2$ (Section~\ref{sec:al_execution}). The per-triangle uncertainty discussed in Appendix~\ref{appendix:per_tri} is converted into a simulate-next decision in this way: the per-triangle field is collapsed to a single number per orientation, every feasible orientation is ranked by that number, and the loop simulates from the high end of the ranking.

\section{Per-triangle prediction and its uncertainty}\label{appendix:per_tri}
\begin{figure}[t]
\centering
\includegraphics[width=0.7\textwidth]{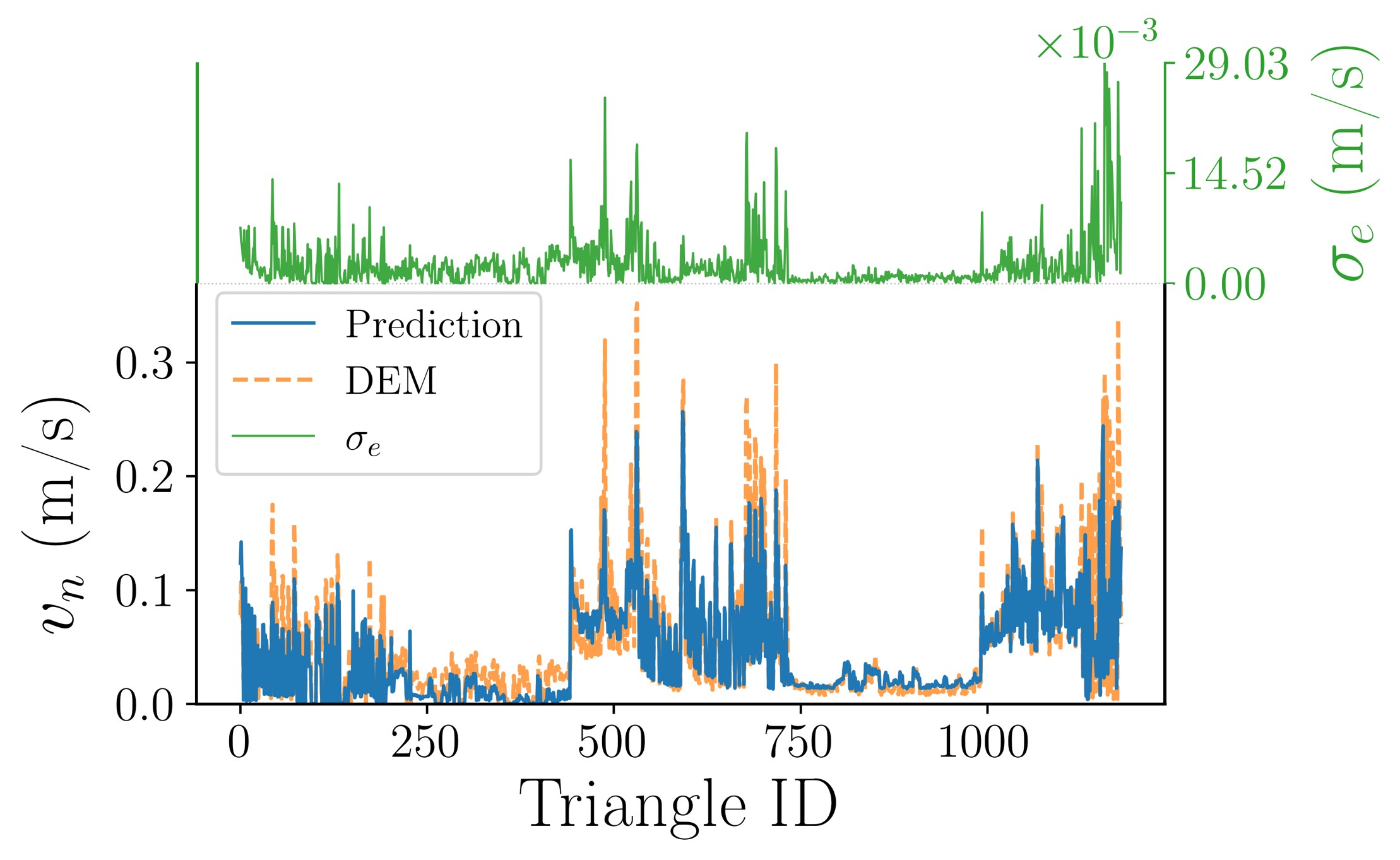}
\caption{Per-triangle prediction and epistemic uncertainty for a representative held-out orientation. The top panel shows the epistemic standard deviation $\sigma_e$ (solid black, right axis); the bottom panel shows the predicted normal impact velocity (solid blue) against the DEM ground truth (dashed red, left axis), both plotted against triangle ID. In the low-velocity, sheltered regions (roughly triangles $200$--$390$ and $790$--$980$) the prediction matches DEM closely and $\sigma_e$ is at its lowest. In the exposed, high-velocity regions the agreement is still good overall but individual peaks are harder to match exactly, and it is in these regions, most visibly beyond triangle $1080$, that $\sigma_e$ rises to its highest values over the surface.}
\label{fig:per_tri_pred}
\end{figure}

Figure~\ref{fig:per_tri_pred} shows the surrogate's per-triangle output for a representative held-out orientation, the predicted normal impact velocity, the DEM ground truth, and the corresponding epistemic uncertainty $\sigma_e$, all plotted against triangle ID. The prediction reproduces the DEM field closely over most of the surface. In the low-velocity regions, roughly triangles $200$--$390$ and $790$--$980$, where the facets are sheltered and see little particle impact, the prediction tracks DEM almost exactly and $\sigma_e$ sits near its lowest values. In the high-velocity regions, triangles $400$--$780$ and $980$--$1179$, where the exposed facets receive the most energetic impacts, the agreement is still good overall. The local peaks are harder to match exactly, however, and the prediction occasionally under- or overshoots a sharp DEM spike, for instance near triangles $580$--$600$ and again past triangle $1080$. These are also the regions where $\sigma_e$ rises, most visibly toward the end of the triangle range, where the uncertainty trace reaches its highest values over the whole surface. The correspondence is therefore qualitative rather than pointwise: the surrogate is not simply more uncertain wherever it is wrong. Rather, the regions of elevated $\sigma_e$ and the regions where the prediction departs from DEM largely coincide, while the flat, low-velocity, low-uncertainty regions are exactly where the prediction is most exact. The same pattern recurs across the other orientations examined: low uncertainty in the sheltered regions, and elevated uncertainty in the exposed, harder-to-predict regions. This recurring pattern motivates aggregating $\sigma_e$ to the orientation level in Appendix~\ref{appendix:sigma_ori}, rather than relying on it triangle by triangle.

%\bibliographystyle{unsrt}
%\bibliography{bibtex}

\end{document}